\documentclass[nohyper,letterpaper,12pt]{JHEP3}

\usepackage{graphicx}
\usepackage{amsmath}

\newcommand{\half}{\frac{1}{2}}
\newcommand{\beq}{\begin{eqnarray}}
\newcommand{\eeq}{\end{eqnarray}}
\newcommand{\nnn}{ \nonumber \\ }
\newcommand{\p}{{\partial}}
\newcommand{\zb}{{\bar{z}}}
\newcommand{\e}{{\epsilon}}
\newcommand{\s}{{\sigma}}
\newcommand{\vev}[1]{{\langle #1 \rangle}}

\newcommand{\ord}[1]{{{\cal O}(#1)}}
\newcommand{\gappeq}{\mathrel{\rlap {\raise.5ex\hbox{$>$}}
{\lower.5ex\hbox{$\sim$}}}}
\newcommand{\lappeq}{\mathrel{\rlap{\raise.5ex\hbox{$<$}}
{\lower.5ex\hbox{$\sim$}}}}
\newcommand{\myref}[1]{(\ref{#1})}

\newcommand{\ben}{\begin{enumerate}}
\newcommand{\een}{\end{enumerate}}

\newcommand{\sqtw}{\sqrt{2}}

\newcommand{\phib}{{\bar \phi}}

\newcommand{\psib}{{\bar \psi}}
\newcommand{\nbf}{{\vec n}}
\newcommand{\mbf}{{\vec m}}
\newcommand{\Fb}{{\bar F}}
\newcommand{\bbar}[1]{{\overline{#1}}}
\newcommand{\chib}{{\bar \chi}}
\newcommand{\fourth}{\frac{1}{4}}
\newcommand{\kbf}{{\vec k}}
\newcommand{\real}{\mathop{{\hbox{Re} \, }}\nolimits}
\newcommand{\imag}{\mathop{{\hbox{Im} \, }}\nolimits}
\newcommand{\Ncal}{{\cal N}}

\def\[{\left [}
\def\]{\right ]}
\def\({\left (}
\def\){\right )}

\preprint{hep-th/0407135 \\ Sept.~10, 2004}

\title{Lattice supersymmetry, superfields and renormalization}

\author{Joel Giedt \\ 
E-mail: \email{giedt@physics.utoronto.ca}}
 
\author{Erich Poppitz \\ 
E-mail: \email{poppitz@physics.utoronto.ca} \\ \\
Department of Physics, University of Toronto \\
60 St. George St., Toronto ON M5S 1A7 Canada}

\abstract{
We study Euclidean lattice formulations of non-gauge supersymmetric models with up to four supercharges
in various dimensions. We formulate the conditions under which the interacting lattice theory can exactly preserve one or more nilpotent anticommuting supersymmetries.  We introduce a superfield formalism, which allows the enumeration of all possible lattice supersymmetry invariants. We use it  to     discuss the formulation of Q-exact 
lattice actions and their renormalization  in a general manner.  In some examples,  one exact  
supersymmetry guarantees finiteness of the continuum limit of the lattice theory.  As a consequence, we   show that the desired quantum continuum limit is obtained without fine tuning for these models. Finally, we discuss the implications and  possible further applications of  our results to the study of gauge and non-gauge models.
}


\begin{document}

\section{Introduction}

\subsection{Motivation}
Supersymmetric field theories enjoy remarkable  perturbative  nonrenormalization
properties, as was first noticed in the 4d Wess-Zumino models
at 1-loop \cite{Wess:1973kz} and then to all orders~\cite{Iliopoulos:1974zv}.
Remarkably, a  nonperturbative  nonrenormalization theorem was
proven for the 4d Wess-Zumino model with a cubic superpotential
interaction \cite{Seiberg:1993vc}.  On the other hand, in
the more interesting and phenomenologically relevant case
of super-QCD, it is known that the tree level superpotential
 does receive nonperturbative corrections \cite{ADS}.
Indeed,   nonperturbative contributions to the
superpotential are often invoked in scenarios for
moduli stabilization and spontaneous supersymmetry breaking
in a variety of supersymmetric extensions to the
Standard Model of particle physics.  

The above examples
illustrate the importance of reliable nonperturbative
methods of analysis in supersymmetric field theories.
While traditional methods based on holomorphy and symmetry, 
including embedding the theories in various string constructions, 
have resulted in stunning progress for many theories, it would be advantageous to have 
a comparable wealth of results from other methods.
The most pressing reason is that we would like
to answer questions that the traditional methods
are not able to address, such as nonperturbative
contributions to nonholomorphic quantities.
One  technique for nonperturbative analysis
which is worth exploring is lattice regularization.
It is this particular approach that we study here.
Moreover, the lattice is the only known fundamental,
nonperturbative definition of a general quantum field theory; 
as a matter of principle, we believe that it is important 
to have such a definition for supersymmetric field theories.

Typically in field theory, one seeks a regulator that
preserves all of the symmetries present at tree level.
Otherwise, symmetry breaking will be induced, producing
spurious results.  In the continuum formulation of
supersymmetric theories, various methods which meet
this requirement have been developed over the years:
a supersymmetric Pauli-Villars sector,
dimensional reduction, supersymmetric higher derivative
terms, etc.  However, in 30 years no fully supersymmetric
lattice regulator of an  interacting supersymmetric field
theory has been constructed.
The best that has been so far achieved are lattice
models where it can be argued that the target supersymmetric
field theory is obtained with little or no fine-tuning in the
quantum continuum limit.  In these cases, the full supersymmetry
is recovered in one of two ways.  First, without fine-tuning, as an accidental
symmetry that results from symmetries of the lattice 
theory \cite{Kaplan:2002wv}-\cite{Sugino:2003yb};
in the case of 4d $\Ncal=1$ super-Yang-Mills,
chiral symmetry guarantees continuum supersymmetry,
either through a modest fine tuning with Wilson fermions \cite{Curci:1986sm}
or, without fine-tuning, through chiral lattice fermions \cite{Kaplan:1999jn}.
The second case is where supersymmetry is recovered due
to finiteness\footnote{
Here, and in the remainder of this article, ``finite'' will
have the following technical meaning:  the sum of diagrams
contributing to any proper vertex has UV degree $D<0$
according to the rules of power-counting in lattice perturbation
theory.  A specific example will be discussed in appendix D.} 
or super-renormalizability 
\cite{Elitzur:1982vh}-\cite{Golterman:1988ta}.
In the constructions considered below, we will see aspects of both  these cases.

\subsection{The problem}
The main difficulty for preserving exact 
supersymmetry on the lattice is to respect the supersymmetry
of the interaction terms.\footnote{For a
lagrangian quadratic in the fields, it is
possible to exactly preserve all supersymmetry on the lattice, 
given  a judicious choice of lattice derivatives; see section \ref{susyqm} and appendix B.
We are not interested in this trivial case, and
do not enumerate the many articles that have
focused solely on such constructions.}  The main obstacle to achieving this is the failure of the Leibnitz rule for
lattice derivatives.\footnote{Not, as sometimes stated, 
the fermion doubling problem; it would only be an obstacle
for chiral supersymmetric theories.}
To elucidate, recall that supersymmetry generators can be represented as differential operators acting on functions of ``superspace" ($x, \theta,... \theta^\prime$). A supersymmetry generator $Q$  typically takes the form:
\beq
\label{generator}
Q = {\partial \over \partial \theta} + \theta \Gamma {\partial \over \partial x}~,
\eeq
where  we omitted various indices (i.e., the details of the constant matrices $\Gamma$, indices of $\theta$, etc.) as inessential for  the general  argument here.
A supersymmetric action is then given as an integral over superspace of a local function of  superfields $\Phi(x, \theta,... \theta^\prime)$:
\beq
\label{lagr1}
S  =  \int d x d \theta ... d  \theta^\prime ~ F(\Phi)~.
\eeq
Supersymmetry  is generated by the $Q$ action on the superfields $\delta_\epsilon \Phi = \epsilon Q \Phi$: 
\beq
\label{vartn1}
\delta_\epsilon S &=& \int d x d \theta ... d \theta^\prime \left[ F(\Phi + \epsilon Q \Phi) - F(\Phi)) \right] \nonumber \\
&=& \int d x d \theta ... d \theta^\prime \;  \epsilon Q F(\Phi) \\
&=&  \int d x d \theta ... d \theta^\prime \; \epsilon 
\left({\partial \over \partial \theta} + \theta  \Gamma {\partial \over \partial x} \right) F (\Phi)~ = 0~. \nonumber
\eeq
The   two terms on the last line vanish separately, for  somewhat different reasons: the first term is zero  because
the $\partial/\partial \theta$ derivative   eliminates the corresponding $\theta$ from the lagrangian and the remaining $\int d \theta$ is zero due to Grassmann integration rules. 

The vanishing of the second term---a total derivative and hence a surface 
term---in the last line of (\ref{vartn1}) is more interesting (from our present 
perspective).  In going from the first to the second line of (\ref{vartn1}), 
we asserted that  $F(\Phi + \epsilon Q \Phi) - F(\Phi)= \epsilon Q F(\Phi)$ and 
implicitly used the Leibnitz rule for spacetime derivatives. On the lattice,  
however, spacetime derivatives are replaced by finite differences, for which 
the Leibnitz rule fails. A naive latticization of a supersymmetric action will 
then have a supersymmetry variation that is a lattice total derivative 
(whose contribution to the variation vanishes, as in (\ref{vartn1})) 
plus corrections of order the lattice spacing, which spoil the supersymmetry 
of the action---a fact that was originally pointed out by Dondi and Nicolai \cite{Dondi:1976tx}.
Thus,  quite generally, a naive latticization
of an interacting supersymmetric lagrangian will lead to a nonsupersymmetric lattice action, whose  supersymmetry variation is proportional to powers of the lattice spacing.

In the naive (i.e., classical) continuum limit, 
supersymmetry is, of course, restored. However,  quantum effects may generate a number of  relevant operators violating the continuum limit supersymmetry---for example, soft masses for scalars, which are not forbidden by any symmetry (but supersymmetry). These then require a possibly large number of fine tunings to recover the supersymmetric continuum limit. The need for fine tuning, while theoretically palatable, renders the lattice studies of interesting nonperturbative phenomena in supersymmetric theories, such as dynamical (super-) symmetry breaking  practically impossible.

\subsection{The approach}
\label{theapproach}
The natural question to ask, then, is  whether it is possible to preserve at least a subset of the supersymmetry transformations in the interacting lattice theory.  As shown above, the source of the difficulty   is the second term in  $Q$ of eqn.~(\ref{generator}), proportional to space-time derivatives. If  a supercharge could be represented, upon conjugation (a change of basis) in a form   involving only $\theta$-derivatives, the  Leibnitz rule for spacetime derivatives would not be needed to ensure supersymmetric invariance of the action. Thus,  the corresponding supersymmetries   stand the chance to be exact symmetries of the lattice interactions.  It is clear---and we will see many examples---that nilpotent anticommuting 
supercharges   can   always be conjugated to pure 
$\theta$-derivatives. We will  explain how a latticization of  the continuum 
action that preserves these nilpotent anticommuting supersymmetries
can  be performed in most cases.  

However, the existence of  nilpotent  supercharges is not sufficient to guarantee that all interactions on the lattice are invariant under the corresponding supersymmetries. 
This will be explained in more detail in  section \ref{20theory}; here we only state our general conclusions in this regard:
\begin{enumerate}
\item If the continuum supersymmetry algebra contains one or more nilpotent 
anticommuting supercharges, supersymmetric actions that are integrals over 
full superspace (e.g., D-terms) can be latticized while preserving all these supercharges.
\item Continuum interaction terms given by integrals over parts of the full superspace 
(e.g., over chiral superspace, F-terms) remain invariant after latticization 
only if there exists a (linear combination of) nilpotent supercharge(s) such that 
the supersymmetry variation of these interactions terms is not a total derivative.
\end{enumerate} 

The above criteria 1.) and 2.)  are necessary    to define a supersymmetric lattice partition function. However, they are not sufficient to ensure that it will have the desired  continuum limit, even classically. As we will see, while studying some 2d $(2,2)$ theories, the   removal of fermion doublers with the simultaneous preservation of the exact supersymmetry of the partition function and of other desirable symmetries of the continuum theory is not always possible within the present approach.

Nevertheless, in many cases we will be able to write down classical lattice actions that preserve part of the continuum supersymmetry and have the desired continuum limit spectrum and interactions. 
One can hope, then, that  the exact lattice supersymmetry,    perhaps  combined with other symmetries of the lattice action,  precludes the generation at the quantum level of relevant operators  that break  the continuum limit supersymmetry. Thus, the goal of this approach to lattice supersymmetry is  to find examples of interacting  lattice theories where reaching the supersymmetric  critical point requires no fine tuning (or, at least, where the number of required fine-tunings is reduced, compared to a naive discretization). Studying this question is the main thrust of this paper. 

In fact,  all current Euclidean lattice formulations  of interacting supersymmetric theories that   
maintain   part of the  continuum supersymmetry  algebra do so by  preserving precisely 
such a nilpotent subset of the supersymmetry transformations.\footnote{The hamiltonian methods using super-DLCQ,  preserve the entire supersymmetry algebra; for a review, see \cite{Lunin:1999ib}.}
The earliest example of a theory preserving part of the continuum limit supersymmetry is the spatial lattice  approach of refs.~\cite{Elitzur:1982vh}--\cite{Sakai:1983dg}, 
yielding  a   lattice hamiltonian (invariant under supersymmetries whose anticommutator does not include lattice translations); there have been numerous studies of this approach 
since, a recent example being \cite{Beccaria:2004ds}.

More recently, there has been a revival of the study of  theories with supersymmetry 
on both  spatial and Euclidean lattices, 
coming from two    different directions. The first is the   
exploitation of relations to topological quantum field 
theory \cite{Catterall:2003wd}, closely related to the earlier work of \cite{Sakai:1983dg}, \cite{Elitzur:1983nj}. 
 The second direction is the construction 
of  supersymmetric noncompact lattice gauge   theories  using a ``top-down" approach, based on the 
orbifolding of ``mother"  models, a.k.a.~``deconstruction" \cite{Kaplan:2002wv}.
Finally, ref.~\cite{Sugino:2003yb} recently made an interesting proposal 
to latticize compact supersymmetric gauge theories, bearing many
features in common with the first of the above  approaches.  
In all these models, the  exact lattice supersymmetries  derive  from one or more nilpotent anticommuting supercharges.

We pause to  mention a few important technical points.  There are cases where the exact lattice supercharges can not be 
conjugated to pure-$\theta$ derivatives. Examples exist both in gauge and non-gauge models with lattice supersymmetry.
A non-gauge example is given in this paper  in our study of  the 2d $(2,2)$ theories
with   twisted mass term, see eqn.~\myref{twistedsusy},    where the two exact nilpotent  lattice 
supercharges do not anticommute---the anticommutator of the  two nilpotent supercharges is proportional to the central charge. 
In the  case of gauge theories (not the topic of this paper), if we insist on avoiding 
 the introduction of unphysical degrees of freedom, we  also expect to have 
exact lattice supersymmetries that anticommute or are nilpotent only up to gauge transformations. 
This is because  the anticommutator of two supercharges in Wess-Zumino ``gauge"
involves gauge transformations, in addition to spacetime translations. 
Appropriate examples are described in the recent literature:  in the ``deconstruction" lattice models with eight 
supercharges of the target theory \cite{Kaplan:2002wv},  anticommutativity of the two exact lattice supercharges 
holds only up to gauge transformations, while in the construction of \cite{Sugino:2003yb} it is  
the nilpotency of the exact supercharge that holds only up to gauge transformations.  
Finally, we also note that while all lattice supersymmetric actions  in this paper  are Q-exact, 
it can happen that terms appear in the action that are only  Q-closed; an example of this can be found 
in the 3d lattice of~\cite{Kaplan:2002wv}.

We summarize this section by stressing the  unifying theme of  all formulations of lattice supersymmetry mentioned above: the  lattice theory exactly  preserves some anticommuting nilpotent (perhaps up to gauge transformations) supersymmetries. 
The rest of the supersymmetries are  approximate  and it is hoped or argued that they are restored in the quantum continuum limit.
In this paper, we   begin to   develop this approach to lattice supersymmetry  in a systematic  ``bottom-up" manner and to explore the possibilities of constructing supersymmetric lattice actions in different models with up to four supercharges in various dimensions.

\subsection{Summary}
Many of the supersymmetric lattice actions found in this paper are not new; 
however, as we will explain in the main text,  they have been written using 
methods different from ours \cite{Sakai:1983dg,Catterall:2003wd}, 
which somewhat obscure the study of symmetries and renormalization. 
We develop a unified  formalism,  which allows for the straightforward
construction of all possible terms invariant under the lattice supersymmetry.
The ability to write down all invariants under the lattice (super)symmetries  
is especially helpful to study the  possible supersymmetric deformations of 
the lattice actions,  their renormalization, the approach to the continuum limit, 
and the  corresponding restoration of (super)symmetries. In addition, we hope 
that our approach can be generalized to gauge models.

We begin, in section \ref{susyqm}, by describing the simplest example of 
a lattice system with one exact supersymmetry: quantum mechanics on a 
Euclidean ``supersymmetric" lattice. All technical elements that are 
important in higher-dimensional examples are present here in their simplest 
form, hence we give sufficient detail. The result is a lattice partition function, 
eqns.~(\ref{QMaction2}) and (\ref{QMZ}),  preserving one exact supersymmetry. 
For this 1d lattice, the full supersymmetry is 
automatically restored in the continuum limit.
In a forthcoming paper, we will describe why this is not
true for, say, the naive discretization.  (Monte Carlo
simulution results presented in \cite{Catterall:2000rv} have already shown
an indication of this.) To hint at the
results which will be shown there, a diagram with UV degree $D=0$
in the lattice perturbation theory gives a finite contribution---coming
from doubler modes associated with the Wilson fermions---that is not
present in the continuum perturbation theory.  This effect is
avoided in the present approach, due to cancellations
that occur due to the exact lattice supersymmetry.
Besides this vital feature, it is worth noting that
the supersymmetric partition function enjoys a number of desirable properties: 
we show that the Witten index can be exactly calculated already at finite lattice 
spacing; furthermore, numerical analysis has shown
that the approach to the supersymmetric continuum 
limit is faster and exhibits boson-fermion spectrum 
degeneracy \cite{Catterall:2000rv,Catterall:2001wx}.  Much of the detail supporting 
these statements is given in appendix A. 

We continue, by increasing the dimensionality of  spacetime, in sections \ref{11susy} 
and \ref{20theory},  where we study the applicability of criteria 1.) and 2.)  
from section \ref{theapproach} to  different  2d and 3d theories with two 
supercharges,   and to 3d and 4d theories with four supercharges. The 
first set of examples, considered in section \ref{11susy},
are the two-supercharge 3d and the $(1,1)$ 2d theories.
These have no nilpotent supercharge that can be preserved by the interactions; 
thus criterion 1.) is already violated.  The second example, considered in section \ref{20theory}, 
is the $(2,0)$ 2d theory.  It illustrates the necessity of criterion 2.). 
This  provides an example  of an important class of theories, where not all 
interaction terms can preserve the nilpotent supersymmetry. Many interesting theories, 
such as the 4d ${\cal{N}}=1$ Wess-Zumino model and its 3d compactification, fall into this class.  

The main body of this paper, section 5,  is devoted to the study of  2d $(2,2)$ models with chiral superfields. We begin, in section \ref{choice}, by examining the two possible choices, ``$A$"--type and ``$B$"--type,  of pairs of supercharges to be preserved by the lattice and explain why we focus on the $A$-type charges. We then  introduce  lattice superfields and the action of  the lattice counterparts of the  continuum limit global symmetries. Much of the technical detail of the construction of lattice superfields is given in appendix B. 
 
In section \ref{22susysigma} we show that the most general lattice action, preserving the $A$-type supercharges and the maximal number of global symmetries, is a lattice version of  the $(2,2)$ sigma model. While this action has the correct naive continuum limit, it suffers a fermion doubling problem. We discuss two possible approaches to the doubling problem, in appendix C and in section \ref{lift2}. The first allows the doublers to be lifted at the expense of breaking all the lattice supersymmetry, while the second preserves one nilpotent supercharge.

In appendix C, we study the twisted mass  version of the sigma model and show that twisted nonlocal mass terms can be used to lift the doublers at the expense of breaking all the lattice supersymmetry. Since generic 2d supersymmetric sigma models are renormalizable (rather than superrenormalizable), a lattice implementation of the models with twisted Wilson terms  may require fine tuning in the quantum continuum limit.\footnote{It  may still be possible that for particular classes of sigma models the fine tuning is minimal or exhausted at one loop; we leave this for future study.} 
 
In section \ref{lift2}, we  find the general formulae for lattice superpotential $F$-terms and show  that  the doublers can be lifted via $F$-type nonlocal mass terms, while preserving  a single nilpotent supercharge.  We show that
the most general supersymmetric lattice action, consistent with the  symmetries, admits non-holomorphic and Lorentz violating relevant terms (which are related to each other by the exact supersymmetry). Thus, the supersymmetric lattice action with the desired $(2,2)$ continuum limit is not generic.  

We then consider renormalization of lattice models whose classical continuum limits
are 2d $(2,2)$ models with good ultraviolet behavior. In appendix D, we study  Wess-Zumino (or Landau-Ginzburg) models   
that are ultraviolet finite and show that  the one exact lattice supersymmetry  
is sufficient to guarantee finiteness of the lattice theory in the zero lattice spacing limit. 
This then allows us,   in section \ref{lift2},  to argue that the non-holomorphic and Lorentz 
violating terms are irrelevant  quantum mechanically and that the $(2,2)$ continuum limit is 
achieved without fine tuning. Further general properties of the supersymmetric lattice version 
of the Wess-Zumino model, such as positivity of the fermion determinant (for square lattices)
are also shown in appendix D.

\subsection{Outlook}
\label{outs}

The examples considered  in this paper show that   
Euclidean actions of supersymmetric  models,  obeying  
the criteria of section \ref{theapproach},    
can be latticized with the simultaneous preservation 
of a number of exact nilpotent supersymmetries, while 
the rest of the supersymmetries are respected only up
to (strictly positive) powers of the lattice spacing. 

To find whether the exact lattice supersymmetry guarantees 
that the  supersymmetric and Lorentz invariant quantum continuum 
limit is achieved without fine tuning requires further analysis. 
In the interesting example of general 2d $(2,2)$ models, we  
showed that the supersymmetric  lattice action with the desired 
continuum limit is not generic. In other words, obtaining the  
$(2,2)$ quantum continuum limit   either involves fine-tuning or 
requires favorable ultraviolet properties of the target continuum 
theories. The 2d $(2,2)$ finite examples,  that are shown here to 
not require fine-tuning, indicate that in 3d and 4d this 
program is more likely to succeed in theories with a large number 
of supersymmetries, since it is these that are typically ultraviolet finite.  

Thus, the scope of the present approach to lattice supersymmetry, 
while limited, includes   other interesting models, whose latticization 
along the lines presented here is worth studying. Some obvious types of 
theories are missing from our analysis, notably, those with more 
supersymmetry, and particularly, gauge theories. It would be 
interesting to make explicit contact between our approach
and the recent proposal of Sugino.  This would allow us to 
investigate, along the lines of this paper, the possible 
deformations and renormalization of the proposed 
supersymmetric $(2,2)$  lattice gauge theory actions.

The results of section \ref{lift2} below (as well as of appendix D) allow us to 
conclude that the $(2,2)$ LG models can be simulated on the lattice.  Depending 
on the superpotential, these models have infrared fixed points described  
by the ${\cal{N}} = 2$ minimal models of conformal field theory (see, e.g., 
the second reference in \cite{Hori:2000kt}). It would be of some interest, at 
the very least as a nonperturbative  check of the lattice techniques, to numerically 
verify the predicted values of critical exponents.  Work is in
progress along these lines, which we intend to report upon in the
near future.

\section{Supersymmetric quantum mechanics on a supersymmetric lattice}
\label{susyqm}
In this section, we consider supersymmetric quantum mechanics on a supersymmetric lattice. This is a simple enough example that illustrates all major techniques we use to construct supersymmetric lattice actions. Thus, in this section, we give sufficient detail. Since the details of the construction  of higher dimensional theories are more notationally involved, many are  given in the appendices. 

The supersymmetry algebra is generated by two real nilpotent supercharges, with  $\{Q_1, Q_2\} = 
2 i  { \partial \over \partial t} $. The supercharges can be represented in terms of differential operators acting on functions of  time $t$ and two real anticommuting variables $(\theta^1, \theta^2)$:
\beq
\label{QMQs}
Q_1 &=& {\partial \over \partial \theta^1} + i \; \theta^2 \; { \partial \over \partial t} = e^{- i \theta^1 \theta^2 { \partial \over \partial t}} \; {\partial \over \partial \theta^1} \; e^{ i \theta^1 \theta^2 { \partial \over \partial t}} \\
Q_2 &=& {\partial \over \partial \theta^2} + i \; \theta^1 \; { \partial \over \partial t} = e^{ i \theta^1 \theta^2 { \partial \over \partial t}} \; {\partial \over \partial \theta^2} \; e^{ - i \theta^1 \theta^2 { \partial \over \partial t}} ~.\nonumber
\eeq
The supercovariant derivatives $D_1, D_2$  anticommute with $Q_1, Q_2$ and are also represented as differential operators:
\beq
D_1 &=& e^{i \theta^1 \theta^2 { \partial \over \partial t}} \; {\partial \over \partial \theta^1} \; e^{ - i \theta^1 \theta^2 { \partial \over \partial t}} \\
D_2 &=& e^{ -i \theta^1 \theta^2 { \partial \over \partial t}} \; {\partial \over \partial \theta^2} \; e^{ i \theta^1 \theta^2 { \partial \over \partial t}} ~,\nonumber
\eeq
obeying $D_1^2 = D_2^2 = 0$, $\{ D_1 , D_2\} = - 2 i {\partial \over \partial t}$.
The theory can be formulated in terms of real superfields, $\Phi$, with the following expansion:
\beq
\Phi (t, \theta^1, \theta^2) = x(t) + \theta^1 \psi(t) + \theta^2 \chi(t) + \theta^1 \theta^2 F(t)~,
\eeq
containing one auxiliary field $F$ and the physical fields $x$, $\psi$, $\chi$.
The continuum action is  a full superspace integral:
\beq
\label{QMcontinuumaction}
S &=& \int d t \; d  \theta^2 d \theta^1 \left( {1 \over 2} D_1 \Phi D_2 \Phi - h(\Phi) \right) \nnn
&=&   \int d t \left( {1 \over 2} \dot{x}^2 + {1 \over 2} F^2 -   i \psi \; \dot \chi - h^\prime(x) F + h^{\prime\prime}(x)\;  \psi \; \chi \right)~, 
\eeq
where primes denote derivatives of $h(x)$ with respect to $x$. After eliminating $F$, the component action reads: 
\beq
S
=    \int d t \left( {1 \over 2} \dot{x}^2 - {1\over 2} h^\prime(x)^2 - i \chi \; \dot \psi - \chi \;  h^{\prime\prime}(x)\;  \psi  \right)
\eeq	
Eqn.~(\ref{QMcontinuumaction}) contains relevant and marginal terms only: $t$ has 
the usual  mass dimension $-1$, and $x(t)$ and $\theta^1, \theta^2$ are assigned 
mass dimensions $-1/2$. The measure $dt \, d \theta^1 \, d \theta^2$ is thus dimensionless as is the 
lagrangian density; the couplings in the superpotential $h(\Phi)$ are all relevant 
because the superfield $\Phi$ has negative mass dimension. 
 
As explained in the introduction, naive discretization of the path integral breaks supersymmetry. Our goal now is to discretize the path integral while preserving one of the two nilpotent supercharges, for example $Q_1$. It follows from (\ref{QMQs}) that, upon conjugation, $Q_1$ can be represented as  a pure  derivative with respect to $\theta^1$:
\beq
\label{conjugation}
Q \equiv {\partial \over \partial \theta^1} = e^{i \theta^1 \theta^2 {\partial \over \partial t}} \; Q_1 \;  e^{ - i \theta^1 \theta^2 {\partial \over \partial t}}~.
\eeq
Similarly, the superfield $\Phi$ in the conjugated  basis is given by:\footnote{To 
the reader familiar with 4d $N=1$ supersymmetry, this conjugation is similar 
to ``going to the chiral basis," where $\bar D_{\dot \alpha}$ and 
$Q_{\alpha}$ are represented by pure $\theta$-derivatives.}
\beq
\label{phiconjugation}
\Phi^\prime(t, \theta^1, \theta^2) &=&  e^{ i \theta^1 \theta^2 {\partial \over \partial t}} \; 
\Phi(t, \theta^1, \theta^2) \; e^{ -i \theta^1 \theta^2 {\partial \over \partial t}} = \Phi(t + i \theta^1 \theta^2, \theta^1, \theta^2) \\
&=& (x + \theta^1 \psi) + \theta^2 ( \chi - \theta^1 ( i \dot{x} + F ) ) ~.
\eeq
The action of $Q$ on the field $\Phi^\prime$ is simply a shift of $\theta^1$. As (\ref{phiconjugation}) shows, the field $\Phi^\prime$ splits into two components, irreducible with respect to the action of  $Q$. 

Since our lattice action will only preserve the $Q$ supersymmetry, from now on we will denote $\theta^1$ simply by $\theta$. Furthermore, we will also continue  $t$ to Euclidean space $t \rightarrow i t$ (to not clutter notation, from now on we will   use $t$ to  denote  Euclidean time). 
In the reduced ``superspace" $(t, \theta)$, the   components of $\Phi^\prime(t)$ of  (\ref{phiconjugation}), which are irreducible under the $Q$ action,  are introduced  as  new superfields and are denoted by $U$ and $\xi$ : 
\beq
\label{superfieldsQ}
U(t) = x(t) + \theta \psi(t), ~~ \xi(t) = \chi(t) - \theta \left(F(t) +  \dot{x}(t)\right) ~.
\eeq
The supersymmetry $Q$  now acts as a purely ``internal" transformation and no change in its action  occurs when 
we discretize time.  Thus, we replace the continuum variable $t$ by a set of 
discrete discrete points $t^i$, $i = 1,\ldots N$, such that $t^i - t^{i - 1} = a$ 
is the lattice spacing; the total size of the Euclidean time circle is 
thus $N a$  ($t^{i+N} \equiv t^i$). We denote $x(t^i)$ simply by $ x^i$, and similarly for the other component fields. We then introduce the discrete version of the superfields (\ref{superfieldsQ}), $U^i$ and $\xi^i$, at every lattice point:
\beq
\label{latticesuperfieldsQ1}
U^i = x^i + \theta \psi^i, ~~ \xi^i = \chi^i - \theta \left(F^i + {x^i - x^{i - 1} \over a} \right)~.
\eeq
In (\ref{latticesuperfieldsQ1}), we chose to discretize $\dot{x}$ using a backward 
lattice derivative. We now see that our superfield $\xi^i$ is slightly nonlocal 
on the lattice (this will be of recurring interest in our higher-dimensional examples). In 1d,  
however,   we can simply change the bosonic lattice variables ($x^i, F^i$) to  new ones, ($x^i, f^i$), defined as follows:
\beq
\label{variablechange}
x^i \rightarrow x^i, ~~ F^i \rightarrow f^i - {x^i - x^{i - 1} \over a}~, ~~~ i = 1, \ldots , N~.
\eeq
 It is easily seen that the Jacobian of this transformation is unity. We then work in terms of the local lattice superfields:
\beq
\label{latticesuperfieldsQ}
U^i = x^i + \theta \psi^i, ~~ \xi^i = \chi^i - \theta f^i ~.
\eeq

The supersymmetry transformations of the lattice component fields $x^i, \psi^i, \chi^i, f^i$, generated by $Q$ are easily read off eqn.~(\ref{latticesuperfieldsQ}), since $\delta_1 \equiv \epsilon^1 Q = \epsilon^1 \partial/\partial \theta$.
The supersymmetry transformation $\delta_2 = \epsilon^2 Q_2$, generated by the second supercharge $Q_2 \equiv \partial/\partial \theta^2 + 2 i \theta \partial_t$ in the conjugate basis (\ref{phiconjugation}),  can also be easily worked out. The result for  $\delta_1$ and $\delta_2$ is:
\beq
\label{secondsusy}
\delta_1 x^i &=& \epsilon^1 \psi^i,~ \delta_1 \psi^i = 0, ~ \delta_1 \chi^i = - \epsilon^1 f^i, ~ \delta_1 f^i  = 0~, \\
\delta_2 x^i &=& \epsilon^2 \chi^i, ~ \delta_2 \psi^i = \epsilon^2 \left(f^i -  {2 \over a}\; \hat{\Delta} x^i\right), ~ \delta_2 \chi^i = 0, ~ \delta_2 f^i = {2\over a}\; \epsilon^2 {\hat{\Delta}} \chi^i~, \nonumber
\eeq
where $\hat{\Delta}$ is a finite difference operator. It can be taken, for example, as the backward difference $\Delta^- x^i= x^{i} - x^{i-1}$ or the symmetric difference  $\Delta^S x^i = (x^{i+1} - x^{i-1})/2$ operator; eqn.~(\ref{latticealgebra}) below holds for any linear finite difference.
It is straightforward to see that the transformations $\delta_1$ and $\delta_2$ on the lattice fields $x^i$, $\psi^i$, $\chi^i$, and $f^i$, obey a discretized form of the continuum supersymmetry algebra:\footnote{In higher dimensional examples, e.g.~4d ${\cal{N}}=1$ supersymmetry and its dimensional reductions, it is similarly straightforward to latticize the algebra (only when acting on linear functions of the fields, as in (\ref{secondsusy}), (\ref{latticealgebra})). The algebra then closes on lattice translations, provided one uses  symmetric differences  to replace   continuum derivatives (to ensure vanishing of various unwanted $\gamma$ matrix combinations in the commutator).}
\beq
\label{latticealgebra}
\[ \delta_1, \delta_2 \] = {2\over a} \; \epsilon^1 \epsilon^2\;  {\hat{\Delta}}~,~~ \delta_1^2 = \delta_2^2 = 0~.
\eeq 

As we argued in the introduction, and will see more explicitly below, only one of the supersymmetries, $\delta_1$, can be exactly preserved by the interactions on the lattice. The variation of the interaction terms under the second supersymmetry $\delta_2$ is of order the lattice spacing, due to the failure of the Leibnitz rule for   
finite difference operators. 

Now we are ready to write the supersymmetric lattice action. We require that the action   be bosonic,  invariant under the $Q$ supersymmetry,  local, and   discrete time-translation invariant. 
Supersymmetry will be respected if our action is an integral over   $\theta$   of a function of the superfields $U^i, \xi^i$ and their $\theta$-derivatives. The mass dimensions of $U^i$ and $\xi^i$ are $-1/2$ and $0$, respectively. 

Consider first the bilinear  candidates for a  local superspace action density. The continuum measure, $\int d t d \theta$, is replaced by the lattice sum $ a \sum_i  \int d\theta$, which  is fermionic and has mass dimension $-1/2$. We are thus interested in 
forming   bilinear fermionic terms of mass dimension $1/2$. 
Denoting by $\Delta$ the backward difference operator, e.g. $\Delta U^i \equiv U^i - U^{i - 1}$,  we find the following complete list of local bilinear marginal and relevant terms: 
\begin{enumerate}
\item $\xi^i \; {\partial \over \partial \theta} \xi^i $ of mass dimension $1/2$ (marginal),
\item $ \xi^i \; {1\over a} \Delta U^i$ of mass dimension $1/2$ (marginal),
\item $ \xi^i \; U^i$ of mass dimension $-1/2$ (relevant),
\item $U^i \;  {\partial \over \partial \theta} U^i $ of mass dimension $-1/2$ (relevant), 
\item ${1\over a} \Delta  U^i \;  {\partial \over \partial \theta} U^i $ of mass dimension $1/2$ (marginal).
\end{enumerate}
For reasons that will become clear shortly, we will   impose  an additional global discrete symmetry:
\beq
\label{symmetry1}
\theta \rightarrow i \theta,  ~~ U^i \rightarrow U^i, ~~ \xi^i \rightarrow i \xi^i~,
\eeq
under which the superspace measure transforms  as $ d\theta \rightarrow - i   d \theta$. The path integral measure $\Pi_i d \psi^i d \chi^i d x^i df^i$ is invariant, as the fermions $\psi$ and $\chi$ transform by $-i$ and $i$, respectively, while the bosons are invariant. 
The first three terms above  get multiplied by $i$ after a symmetry transformation, as required by invariance of the action.  The fourth and fifth term get multiplied by $- i$  and the corresponding terms in the action are not  invariant.
Thus, the most general bilinear local action consistent with the imposed symmetries is given by: 
\beq
\label{QMaction}
S = - \sum_i  a \int  d \theta \left(  {1 \over 2} \;  \xi^i \; {\partial \over \partial \theta} \xi^i  + \xi^i \; {1\over a} \Delta U^i + m \;   \xi^i \; U^i\right)~,
\eeq
where we have introduced a mass parameter $m$ consistent with the dimension
of the relevant term (it is easily seen that we have obtained a latticized 
form of the action of the  supersymmetric oscillator);
furthermore, note that rescaling of the superfields and $m$ always
permits one to bring the action to the form \myref{QMaction}.

Let us now study the variation  of the quadratic action, eqn.~(\ref{QMaction}),  under the second supersymmetry  $\delta_2$ (\ref{secondsusy}). It is easy to see that the free action is invariant under both $\delta_1$ and $\delta_2$,  provided   $\hat{\Delta}$ in  (\ref{secondsusy}) is taken to be the symmetric difference operator. 
Note that in the action, we use instead 
the backward derivative in order to avoid the fermion and boson doubling which appears whenever a symmetric derivative is used. Using the backward 
finite difference  corresponds to adding a supersymmetric Wilson term with a particular value of $r$, as follows from $\Delta^{-} = \Delta^{S} - \half \Delta^2$ (where $\Delta^2 x^i = x^{i+1} + x^{i-1} - 2 x^i$ is the lattice laplacian from the Wilson term). Any value of $r$ is consistent with both $\delta_1$ and $\delta_2$ invariances.

Introducing a superpotential interaction $h(x)$ in (\ref{QMaction}) is now trivial: one simply replaces  $m U^i \rightarrow h^\prime(U^i)$ in the last term:
\beq
\label{QMaction1}
S =  - \sum_i  a \int  d \theta \left(   {1 \over 2} \; \xi^i \; {\partial \over \partial \theta} \xi^i  + \xi^i \; {1\over a} \Delta U^i +   \xi^i \; h^\prime(U^i) \right)~.
\eeq
After eliminating the auxiliary field $f^i$ via its equation of motion, the component action becomes: 
\beq
\label{QMaction2}
S = \sum\limits_{i=1}^N a \left(  {1\over 2}  \left(  h^\prime(x^i) + {x^i -  x^{i-1} \over a} \right)^2  +    \chi^i   h^{\prime \prime} (x^i) \psi^i + \chi^i \; {\psi^i - \psi^{i - 1} \over a}  \right)~.
\eeq
This is nothing other than the action introduced in \cite{Catterall:2000rv},
denoted there as ``$S_{new}$.''  The action was also studied in \cite{Catterall:2001wx},
and related to a topological field theory construction in \cite{Catterall:2003wd}.

The action (\ref{QMaction2}) can now be used to define the following supersymmetric lattice partition function:
\beq
\label{QMZ}
Z_N =   - c^N \prod\limits_{i=1}^N \int d\chi^i  d \psi^i \int\limits_{- \infty}^{\infty} d x^i \;  e^{- S}~,
\eeq
where $c^N$ a normalization constant, $c = (2 \pi a)^{-1/2}$, and a minus sign is included for convenience (see appendix A).
The supersymmetric lattice partition function $Z_N$ has a number of nice properties, discussed  in detail in appendix A. We enumerate the  results below, along with some comments:
\begin{enumerate}
\item  Despite the fact that  the action (\ref{QMaction2}) is not reflection positive, in the small-$a$ limit $Z_N$ defines a hermitian hamiltonian,  $H_{SQM} =({\hat{p}^2}  + {h^\prime(\hat{q})^2 }  -    {h^{\prime\prime} (\hat{q}) } [ \hat{b}^\dagger , \hat{b} ])/2$, acting in a positive norm Hilbert space. We show this using the transfer matrix formalism, see eqns.~(\ref{Tmatrix}--\ref{Tmatrix5}).
 Thus, in the small lattice spacing limit, $Z_N$ approaches the Witten index of the supersymmetric quantum mechanics:
 \beq
\lim_{N \rightarrow \infty, a \rightarrow 0} Z_N = 
{\rm tr}\; (-1)^F e^{- \beta H_{SQM}}~, ~~~~\beta \equiv N a~-~ {\rm fixed}~.
\eeq 
\item Moreover, we also show that  the exact  supersymmetry of the partition function (\ref{QMZ}) can be used to exactly compute $Z_N$ for any superpotential and for any value of $N$,  $a$. We show that $Z_N$  coincides with the continuum theory Witten index  already at 
finite lattice spacing, see eqns.~(\ref{WI1}--\ref{WI2}).
\item The transfer matrix representation can also be used to give a lattice formulation 
of the finite-temperature partition function $Z_{N}(\beta) = {\rm tr}  \; e^{- \beta H_{SQM}}$. 
In terms of the
lattice action and functional measure, this amounts to
antiperiodic boundary conditions for the fermions.
The exact  supersymmetry of $Z_N(\beta)$ is now broken globally by the boundary conditions  but is locally present and recovered in the low temperature limit. (In contrast to $Z_N$, $Z_N(\beta)$ is generally not exactly calculable.) 
\item Another consequence of the exact supersymmetry of the partition function 
is the following.  Numerical simulations of correlation functions 
using the supersymmetric partition function $Z_N$ have shown \cite{Catterall:2000rv} 
convergence to the fully supersymmetric continuum limit; by contrast,
similar simulations using a naive nonsupersymmetric discretization do not.
In the case of the supersymmetric $Z_N$, the spectrum is degenerate at
any value of $a$ and $N$ and some Ward identities associated with
the nonexact supersymmetry have been observed to hold within
numerical error \cite{Catterall:2000rv,Catterall:2001wx}.
\end{enumerate}

We pause to note that had we added the  $ U^{i-1}  {\partial \over \partial \theta} U^i$ term (the only nonvanishing bilinear term   forbidden by the extra discrete symmetry (\ref{symmetry1})), which gives rise to an extra $\psi^i \psi^{i - 1}$ term in the component action (\ref{QMaction2}), we would have also obtained a supersymmetric discretized partition function. One can show, however, using the transfer matrix formalism, that its continuum limit  does not  define  a hermitean  hamiltonian system.

One  final comment, which
has already been emphasized in \cite{Catterall:2000rv,Catterall:2001wx,Catterall:2003wd},
is that if we introduce the  new ``Nicolai variable:"
\beq
\label{Nicolai1}
{\cal{N}}^i(x) \equiv  h^\prime(x^i) + {x^i - x^{i-1}\over a}~,
\eeq
we can   rewrite (\ref{QMaction2}) in the following form:
\beq
\label{QMaction3}
S = \sum\limits_{i=1}^N a  \left( {1\over 2} {\cal{N}}^i (x)^2+ \chi^i \; {\partial {\cal{N}}^i(x) \over \partial x^j } \; \psi^j \right)~,
\eeq
 where a sum over repeated indices in the fermion term is understood.
This form   of the supersymmetric lattice action for supersymmetric 
quantum mechanics has been obtained before by discretization of the continuum Nicolai 
variables. We note, however, that  our derivation of the supersymmetric lattice action
assumed no prior knowledge of the existence of a local Nicolai map. Thus, we expect 
that our ``lattice superfield" approach is more general and can be applied also to systems 
where a local Nicolai map is not known, for example quantum mechanics on non-flat manifolds; 
we leave this exercise for future study.

We now continue with the application of our formalism to higher dimensional models.

\section{${\bf (1,1)}$ 2d  and  ${\bf {\cal{N}}=1}$ 3d supersymmetry}
\label{11susy}
The next step is to  consider higher dimensional theories also with two supercharges. Such Lorentz invariant theories exist in two and three dimensions; $(1,0)$ theories also exist in 2d but have no nilpotent supercharge. 
In this section, we review arguments (for completeness) demonstrating
that in $(1,1)$ 2d  and  minimal (${\cal{N}}=1$) 3d supersymmetry 
a nilpotent supercharge does not exist. 
Thus, $Q$-supersymmetry as a path to exact lattice supersymmetry 
is not an option in these cases. Nevertheless, the number of fine-tunings 
required in a nonsupersymmetric lattice theory may be small, or not needed at all,   
due to the super-renormalizability of  some of these theories, see  \cite{Golterman:1988ta}. We will not  consider these theories further, as our  focus here is on supersymmetric lattice actions. 

The $(1,1)$ 2d algebra can be written
($\alpha,\beta,\mu = 1,2$) as:
\beq
\{ Q_\alpha, Q_\beta \} = 2i \gamma_{\alpha \beta}^\mu \p_\mu~.
\eeq
Symmetry of the l.h.s. requires that $\gamma^\mu$ be symmetric.
Without loss of generality we may choose $\gamma^1 = \s_1$.
It follows from the symmetry requirement and the Euclidean Clifford
algebra that (up to a sign) $\gamma_2=\s_3$.  Neither
supercharge is nilpotent.  Suppose we form the most general
linear combination:
\beq
Q = c_1 Q_1 + c_2 Q_2~.
\eeq
Then it follows that:
\beq
Q^2 = 2i c_1 c_2 \p_1 + i(c_1^2 - c_2^2) \p_2~.
\eeq
Clearly $Q^2=0$ has only the trivial solution $c_1=c_2=0$.
The case of Lorentzian metric is simply obtained by the
identifications $\gamma_2=i\gamma_0$ and $\p_2 = -i \p_0$.
The impossibility of nontrivial nilpotent supercharge is
unchanged.

In 3d the ${\cal{N}} = 1$  algebra can be written
($\alpha,\beta = 1,2$ and $\mu=1,2,3$):
\beq
\{ Q_\alpha, Q_\beta \} = 2i (\e \gamma^\mu)_{\alpha \beta} \p_\mu~,
\eeq
where $\e=i\s_2$. 
Symmetry of the l.h.s. requires that $\e \gamma^\mu$ be
symmetric.  Without loss of generality this condition and
the Euclidean Clifford algebra for $\gamma^\mu$ can be satisfied
by choosing $\gamma^\mu = \s_\mu$.  Once again neither
of the supercharges are nilpotent.  Furthermore we find that:
\beq
Q^2 = i(c_1^2 - c_2^2) \p_1 - (c_1^2 + c_2^2) \p_2 - 2i c_1 c_2 \p_3~.
\eeq
Again, $Q^2=0$ has only the trivial solution $c_1=c_2=0$.
For a Lorentzian metric, one simply takes
$\gamma_3=i\gamma_0$ and $\p_3 = -i \p_0$,
to arrive at the  conclusion that there is no nilpotent supercharge in this case as well.

\section{${\bf (2,0)}$ 2d theory and related 3d and 4d theories}
\label{20theory}
Continuing our survey of supersymmetric theories, we note that another theory with only two supercharges exists only in 2d---the $(2,0)$ theory. It provides an important example 
of a theory where nilpotent supercharges exist, i.e. a theory that obeys criterion 1.) of the introduction,
but violates criterion 2.). Hence, not all continuum interactions preserve the nilpotent supersymmetry on the lattice. A similar conclusion also holds for the 4d ${\cal{N}} = 1$ Wess-Zumino model and for its  compactification to 3d.

The $(2,0)$ algebra (here and in the following sections, our notation for 2d supersymmetry is as in, e.g. \cite{Hori:2000kt}) is generated by two supercharges $Q_+$ and $\bar{Q}_+$, obeying:
\beq
\label{20susyalgebra}
\left\{ Q_{+}, \bar{Q}_+ \right\} = - 2 i \partial_+~, ~~ Q_+^2 = \bar{Q}_+^2 = 0~, ~~ \partial_\pm = {1 \over 2} ( \partial_0 \pm \partial_1)~.
\eeq
The supercharges and covariant derivatives can be represented by differential operators acting on the $(x^{\pm}, \theta^+, \bar\theta^+)$ superspace as: 
\beq
\label{20generators}
Q_+ &=& {\partial \over \partial \theta^+} + i \bar\theta^+ \partial_+ ~, ~~\bar{Q}_+ = - {\partial \over \partial \bar\theta^+} - i \theta^+ \partial_+\nonumber \\
D_+ &=& {\partial \over \partial \theta^+} - i \bar\theta^+ \partial_+ ~, ~~ \bar{D}_+ = - {\partial \over \partial \bar\theta^+} + i \theta^+ \partial_+~. 
\eeq
We will consider, as in all cases in this paper, theories of scalar and fermion fields. In the continuum $(2,0)$ theories, these fall into chiral scalar, $\Phi$, and chiral fermion, $\Psi_-$, multiplets. Chirality means, as usual,  that they are subject to the supersymmetric constraint: $\bar{D}_+ \Phi = 0$, $\bar{D}_+ \Psi_- = 0$ (the complex conjugates of $\Phi$ and $\Psi_-$ are antichiral). The chiral scalar and chiral fermion multiplets have the following component expansions (and similar for their antichiral complex conjugates):
\beq
\label{20fields}
\Phi &=& \phi + \theta^+ \psi_+ - i \theta^+ \bar\theta^+ \partial_+ \phi \\
\Psi_- &=& \psi_- + \theta^+ G  - i \theta^+ \bar\theta^+ \partial_+ \psi_- ~,\nonumber 
\eeq
where $\phi$ is a complex scalar field, $\psi_{\pm}$---physical fermion fields and $G$---an auxiliary field. The continuum invariant actions are given in terms of ``$D$" and ``$F$" terms:
\beq
\label{20action}
L_D &=& \int d^2 x \int d \theta^+ d\bar\theta^+  \left( i \bar\Phi \partial_- \Phi + \bar\Psi_- \Psi_- \right) \nonumber \\
L_F &=& \int d^2 x \int d \theta^+ \Psi_- V(\Phi) \bigg\vert_{\bar\theta^+ = 0}~  \\
L_{\bar{F}} &=& \int d^2 x \int d \bar\theta^+ \bar\Psi_- V(\bar\Phi) \bigg\vert_{\theta^+ = 0}~ \nonumber
\eeq
Our latticization of (\ref{20action}) will proceed in complete analogy with the quantum mechanical example and so we omit many of the tedious steps. 

We   choose to preserve one of the nilpotent generators on the lattice, say $Q_+$. Just as in the quantum mechanics case, we transform  to a basis where $Q_+$ is given by $\partial/\partial \theta^+$ and to Euclidean signature, by defining $x^0 = - i x^2$, $z = x^1 + i x^2$,  $\partial_+ = \partial_{\bar z}$, $ \partial_- =  - \partial_z$. We also replace the continuum by a 2d square lattice and the derivatives by appropriate 
finite differences. 
Thus, we find that the superfields $\Phi, \bar\Phi, \Psi_-$, and $\bar\Psi_-$ decompose into  the following ``lattice superfield" components, irreducible under the $Q_+$ action: 
\beq 
\label{20latticefields}
\Phi &\rightarrow&  \phi + \theta^+ \psi_+ ,   \\
\bar\Phi & \rightarrow & \phib, \nonumber\\
& &
 \psib_+ - 2 i \theta^+ \Delta_{\bar z} \phib \nonumber \\
\Psi_- &\rightarrow& \psi_- + \theta^+ G , \nonumber \\
\bar\Psi_- &\rightarrow& \psib_-,\nonumber \\ && \bar{G} - 2 i \theta^+ \Delta_{\bar z} \psib_- \nonumber ,
\eeq
where $\Delta_{\bar{z}}$ is a discretization of $\partial_{\bar z}$ (every line on the r.h.s. represents a superfield irreducible under the $Q_+$ action). 
The supersymmetry transformations of the component fields are immediately read off eqn.~(\ref{20latticefields}), as the $Q_+$ transformations are simple shifts in $\theta^+$; thus, for example, $\phib$ is $Q_+$ invariant, while $\delta_{\epsilon_+} \psib_+  = - 2 i \epsilon_+ \Delta_z \phib$, etc.

Using these superfields, it is straightforward to see that the $L_D$ and $L_F$ terms can be written on the lattice in a form that preserves the $Q_+$ supersymmetry, i.e. shifts in $\theta^+$; we do not give the details since this is not our main point here. 

More important  is the fact that the $L_{\bar{F}}$ interaction  on the lattice can not be written in a $Q_+$ invariant form.  This is because $L_{\bar F}$ is written as an integral over the antichiral $(2,0)$ superspace and its $Q_+$ variation requires use of the Leibnitz rule. To see this, notice that from (\ref{20action}), we find, in the continuum: 
\beq
\label{lbarf}
L_{\bar F} = \int d^2 x \left(\;  \bar{G} V(\phib)  + \psib_- \psib_+ V^\prime(\phib)   \; \right) ~.
\eeq
Under $Q_+$ transformations  $\phib$ and $\psib_-$ are inert, see (\ref{20latticefields}); however, $\psib_+$ and $\bar G$ shift into derivatives of $\phib$ and $\psib_-$, respectively. Thus, the  $Q_+$ variation of  $L_{\bar F}$ is  proportional to 
$\partial_{\bar z} \psib_- V(\phib) + \psib_- V^\prime(\phib) \partial_{\bar z} \phib = \partial_{\bar z} \left( \psib_- V(\phib) \right)$, and hence is a total derivative. Thus,  the invariance of the antichiral superspace integral, even under the nilpotent $Q_+$, requires  the validity of 
the Leibnitz rule. Hence we can not  latticize the action (\ref{20action}) in a manner that preserves supersymmetry of   the antichiral interaction 
terms.\footnote{One ``loophole," which we leave for future work is the construction of  $Q_+$-invariant  $(2,0)$ sigma model lattice actions (they only have $D$-type interactions, so the ``${\bar{F}}$-term  obstruction" does not apply) and the study of related  issues of doublers and anomalies, which are likely to  arise, as in $(2,2)$ sigma models.}

The situation with the $\bar{F}$ terms the $(2,0)$ theory is repeated in a number of  other two-, three-, and four-dimensional models, where nilpotent supercharges exist---the ``B"-choice of lattice supercharges in $(2,2)$ 2d theories (see following section and appendix B),  the ${\cal{N}} = 2$ 3d theory, and the ${\cal{N}} = 1$ 4d Wess-Zumino model. In all these four-supercharge cases, in standard notation,  two anticommuting nilpotent supercharges exist and can be chosen to be, say, the $Q_\alpha$ ($\alpha = 1,2$ is the $SL(2,C)$ index). Similar to our discussion above, the antichiral part of the interaction lagrangian in these models can not be latticized in a manner preserving the nilpotent supercharges, since vanishing of  its $Q_\alpha$ variation requires validity of the Leibnitz rule.

\section{${\bf (2,2)}$ 2d supersymmetry: the $A$-type supersymmetric  lattice}
\label{22section}
\subsection{Nilpotent charges, superfields, and global symmetries}
\label{choice}
We are thus led to consider theories with four supercharges in two dimensions, which will be the main focus of this paper.
The $(2,2)$  2d supersymmetry algebra with no central charges,\footnote{For a lattice realization of a theory with nonzero central charges, see appendix C.} in Minkowski space,  is simply the dimensional reduction of the ${\cal{N}}=1$ supersymmetry algebra in four dimensions:
\beq
\label{22susyalgebra}
&& \left\{ Q_{\pm}, \bar{Q}_{\pm} \right\} = - 2 i \partial_{\pm}~, 
~~ Q_{\pm}^2 = \bar{Q}_{\pm}^2 = 0~, ~~ Q_{\pm}^\dagger = \bar{Q}_{\pm}~, ~~ \left\{ Q_{+}, Q_{-} \right\} = 0~, \nnn
&& ~~ \partial_\pm = {1 \over 2} ( \partial_0 \pm \partial_1)~.
\eeq
As explained in the introduction, if we were to construct a  lattice action of an interacting supersymmetric theory,  the best we can hope for is to explicitly preserve  only a subset of the four supercharges---those, whose anticommutators do not involve translations. Clearly, from eqn.~(\ref{22susyalgebra}), in the $(2,2)$ case, 
 the maximal number of nilpotent anticommuting supersymmetry generators is two. 

There are two, up to hermitean conjugation,  possible choices of nilpotent anticommuting generators. We will call them   ``$A$-type," taking the $\bar{Q}_+$, $Q_-$ pair of generators,
and  ``$B$-type," taking the  $\bar{Q}_+, \bar{Q}_-$   pair of supercharges.  The $B$-type choice is also possible in the 3d ${\cal{N}}=2$ and 4d ${\cal{N}}=1$ theories; the difficulties   that this choice of exact lattice supersymmetries faces were already discussed in the previous section. 

Thus, from now on we focus on the $A$ choice. This choice is unique to 2d, since $Q_-$ and $\bar{Q}_+$ cease to  anticommute when the algebra (\ref{22susyalgebra}) is uplifted to 3d and 4d---their anticommutator involves a translation in the ``extra" spatial directions. Thus, in 3d and 4d they can not be simultaneously brought into a form involving only $\theta$ derivatives.

We will study theories, whose continuum limit are  theories of chiral superfields only:  
$(2,2)$ sigma models or Landau-Ginzburg (LG) models. 
To address the task of writing the most general action consistent with 
the exact lattice supersymmetry, we will develop a ``lattice superfield" formalism, 
making the study of invariant interactions and their symmetries extremely straightforward. 
We follow  the same steps as in the quantum mechanics case. Since, conceptually, there are 
no new steps involved, but the notation is significantly messier,
we give  all details in appendix B. 

Just as in the  quantum mechanics case of section \ref{susyqm}, the end result of the ``lattice superfield" construction  is the introduction of  the following   $A$-type lattice superfields, which correspond to the irreducible (under the $Q_-$ and $\bar{Q}_+$ action) components of continuum chiral superfields: 
\beq
\label{latticesuperfields}
U_\mbf &=& \phi_\mbf+ \theta^- \psi_{-,\mbf}~, \nonumber \\
\Xi_\mbf&=& \nonumber \psi_{+,\mbf} - \bar\theta^+ i \Delta_{\bar z} \phi_\mbf 
	+ \theta^- F_\mbf +  i  \theta^- \bar\theta^+  \Delta_{\bar z} \psi_{-,\mbf}~ , \nonumber \\
{\bar{U}}_\mbf&=& \phib_\mbf - \bar\theta^+ \psib_{+,\mbf}~, \\
{\bar\Xi}_\mbf &=& \psib_{-,\mbf} - \theta^- i \Delta_z \phib_\mbf
-  \bar\theta^+ \Fb_\mbf + i  \theta^-   \bar\theta^+  \Delta_z \psib_{+,\mbf}~.  \nonumber 
\eeq
We note that the fermionic lattice superfields $\Xi$ and $\bar\Xi$ are slightly nonlocal on the lattice, because of the appearance of 
finite difference operators,  and that the fields $U, \Xi$  are not independent, as $\Xi = \psi_+ + \theta^- F - \bar\theta^+ i \Delta_{\bar z} U$. This is because  $U$ and $\Xi$ are   the components of a continuum chiral superfield in the appropriate basis. As explained in   Appendix B, we denote $\Delta_z \equiv \Delta_1 - i \Delta_2$, and $\Delta_{\bar z} \equiv \Delta_1 + i \Delta_2$, where $\Delta_i$ is a   
finite difference operator in the $i$-th direction. As in  quantum mechanics, the supersymmetry transformations of the lattice superfields under  $Q_-$, $\bar{Q}_+$ are simply read off eqn.~(\ref{latticesuperfields}), since the corresponding  supersymmetries are now simply  shifts of $\theta^-$ and $\bar\theta^+$.

Any function of the lattice superfields (\ref{latticesuperfields}) integrated over the $\theta^-$, $\bar\theta^+$ superspace is then trivially invariant under the 
two supersymmetries. However, as we will see shortly (in section \ref{susyqm}, we already saw an example  in the simpler case of quantum mechanics),  many of the superspace lattice invariants do  not correspond to  continuum $(2,2)$ supersymmetric  and Euclidean rotation invariant actions.
This should not come as a surprise, as the preservation of only the $Q_-$ and $\bar{Q}_+$ supersymmetries  is not restrictive enough to recover all  continuum limit symmetries. 
In order to achieve the desired $(2,2)$-supersymmetric continuum limit,  we  have to  impose additional symmetries on the lattice actions.

To this end, recall that     $(2,2)$ supersymmetric continuum  theories have  three classical global $U(1)$ symmetries:  Euclidean rotation, $U(1)_{E}$, as well as the vector and axial $U(1)_{V/A}$ transformations, whose action is:
\beq
\label{U1s}
U(1)_V: &~&~ \psi_{\pm} \rightarrow e^{ i \beta} \psi_{\pm}~,~~ \psib_{\pm} \rightarrow  e^{- i \beta} \psib_{\pm}~, ~~ F \rightarrow e^{2 i \beta} F~, ~~\bar{F} \rightarrow e^{-2 i \beta} \bar{F}~\\
U(1)_A: &~&~ \psi_{\pm} \rightarrow e^{\pm i \omega} \psi_{\pm}~,~~ \psib_{\pm} \rightarrow  e^{\mp i \omega} \psib_{\pm} \nonumber ~,
\eeq
and trivial on other fields.
The   vector and axial transformations above are obtained by assigning zero $V$ and $A$ charge to the continuum superfields, as is indicated by the trivial action on the scalar components in (\ref{U1s}). 
The continuum $U(1)_E$ is, of course, broken on the lattice.
On a square lattice, a discrete $Z_4$ subgroup can be preserved, with action, denoting by $\Phi$ any of the bosonic fields,  given by: 
\beq
\label{rotation}
Z_4: &~& \Phi_{m_1, m_2} \rightarrow  \Phi_{m_2, - m_1} ,\nonumber \\
~ &~&  \psi_{\pm,m_1, m_2}  \rightarrow  e^{\mp {i \pi\over 4}} 
\; \psi_{\pm,m_2, - m_1} , \\
~&~& 
{\psib}_{\pm,m_1, m_2}  \rightarrow   e^{\mp {i \pi\over 4}} \; \psib_{\pm,m_2, - m_1} ~. \nonumber 
\eeq 
Two additional symmetries will be important for us. Fermion parity is defined in the  usual manner:
\beq
\label{fermionparity}
Z_{2 F}: \theta^- \rightarrow - \theta^-,~ ~\bar\theta^+ \rightarrow - \bar\theta^+,~~ \Xi \rightarrow - \Xi, ~~ \bar\Xi \rightarrow - \bar\Xi,~ ~U \rightarrow U, ~ ~\bar{U} \rightarrow \bar{U}~.
\eeq
Finally, we define an involution (hereafter called  $I$)  transformation on the lattice action, which  
includes complex conjugation of the parameters in the action and c-numbers in
superfields, a reversal of the order of fermionic  fields,  and  subsequent  replacements of all lattice component fields as follows: 
\beq
\label{involution}
 I &: &\theta^- \rightarrow - i   \bar\theta^+ , ~\bar\theta^+ \rightarrow  - i  \theta^-, ~\psi_\pm \rightarrow i \psib_\mp , ~ \psib_\pm \rightarrow i  \psi_\mp,   \nnn
&& ~ \phi \rightarrow \phib,  ~ \phib \rightarrow \phi,~ F\rightarrow \bar{F}, ~ \bar{F} \rightarrow F~. 
\eeq
The action of these symmetries on the lattice superfields (\ref{latticesuperfields}) is summarized below for further use:
\beq
\label{superfieldsymmetries}
U(1)_V&:&   \theta^- \rightarrow e^{-i \beta} \; \theta^-,~~\bar\theta^+ \rightarrow e^{i \beta}\; 
\bar\theta^+,~~\Xi \rightarrow e^{i \beta} \; \Xi,~~ \bar\Xi \rightarrow e^{- i \beta} \; \bar\Xi, \nnn
&& ~~U \rightarrow U,~~  \bar{U} \rightarrow \bar{U}~,      \nonumber  \\
U(1)_A&:&  \theta^- \rightarrow e^{i \omega} \; \theta^-,~~ \bar\theta^+ \rightarrow e^{i \omega}\; \bar\theta^+,~~ \Xi \rightarrow e^{i \omega}\; \Xi,~~ \bar\Xi\rightarrow e^{i \omega}\; \Xi , \nnn
&& ~~ U \rightarrow U, ~~ \bar{U} \rightarrow \bar{U}~, 
\eeq
\beq
Z_4 &:& \theta^- \rightarrow e^{ - i {\pi \over 4}} \theta^-,~ 
\bar\theta^+ \rightarrow e^{i {\pi \over 4}} \bar\theta^+, 
~\Xi_{m_1 , m_2} \rightarrow e^{ - i {\pi \over 4}} \Xi_{m_2, - m_1}, \nnn
&& ~\bar\Xi_{m_1,m_2} \rightarrow e^{  i {\pi \over 4}}\bar\Xi_{m_2, - m_1}, 
~~ U_{m_1, m_2} \rightarrow U_{m_2, - m_1} ,~ \nnn
&& {\bar{U}}_{m_1, m_2} \rightarrow {\bar{U}}_{m_2, -m_1}~. \nonumber \\
Z_{2 F}&:&  \theta^- \rightarrow - \theta^-,~ ~\bar\theta^+ \rightarrow - \bar\theta^+,~~ \Xi \rightarrow - \Xi, ~~ \bar\Xi \rightarrow - \bar\Xi,~\nnn
&&  ~U \rightarrow U, ~ ~\bar{U} \rightarrow \bar{U}~, \nonumber \\
I &:&  \Xi \rightarrow  i  \; \bar\Xi,~~ \bar\Xi \rightarrow i \;\Xi, ~~U \rightarrow \bar{U}, ~~\bar{U} \rightarrow U, ~~ d\theta^- d \bar\theta^+ \rightarrow  - d\theta^- d \bar\theta^+~. \nonumber
\eeq
In the next section, we will use these symmetries to restrict the possible terms that can appear in the action. We stress again that the transformation properties of the lattice superfields under the involution $I$ hold for any definition of the lattice derivatives, because the involution interchanges $\psi_+$  and $\psib_-$ (and $\psi_-$ with $\psib_+$). In this paper, we will take them to be the symmetric difference operators.

\subsection{The   ${\bf (2,2)}$ chiral sigma model: supersymmetric lattice action}
\label{22susysigma}
We now turn   to the  lattice theory  of chiral superfields, preserving the $A$-type supercharges, 
by classifying the  terms allowed by the lattice supersymmetry and other imposed global symmetries.

Consider first  superspace invariants made only of $U$ and $\bar{U}$.
The simplest ones are half-superspace integrals of functions of $U$ or $\bar{U}$. The terms $\int d \theta^- V (U)$ as well as $\int d \bar\theta^+ V^*(\bar{U})$ are clearly supersymmetric, but violate fermion parity. 
The next simplest possibility is to  allow for a full superspace integral (and also allow for superspace derivatives of the superfields to appear in the action) and consider  lattice supersymmetry invariants of the form:
\beq
\label{Uterms}
\int d \theta^- d \bar\theta^+ f(U, \bar{U},  {\partial U \over \partial \theta^-} , {\partial\bar{U}\over \partial \bar\theta^+} )~,
\eeq
where we do not indicate the dependence on the lattice points (allowing, at this point, for  generic local or nonlocal interactions). Interactions like (\ref{Uterms})  can be arranged to preserve  the $U(1)_V$, $Z_{2 F}$, and $I$ (if appropriate complex conjugation conditions are imposed on $f$)  symmetries; they can also be made  $Z_4$ invariant. Local (or nonlocal) terms like (\ref{Uterms}) can thus only be forbidden by imposing the $U(1)_A$   as no interaction of the above form preserves the axial symmetry.  

The $U(1)_A$ requires that any full superspace invariant be quadratic in  $\Xi$ or $\bar\Xi$, as these are the only fields that have axial charges opposite that of the superspace measure.  Thus, we 
now  allow the possibility of having also $\Xi$ and $\bar\Xi$ appear in the lattice action. Consider the most general $U(1)_V$, $U(1)_A$, $Z_{2 F}$, $I$, and $Z_4$ invariant  lattice superfield action. Also, it suffices to  consider only invariants that do not involve $\theta$-derivatives of either $U, \bar{U}$ or $\Xi$, $\bar\Xi$;  by dimensional analysis, such interactions correspond to irrelevant  terms in the continuum limit.
The most general lattice action with these symmetries, which is a {\it local} function of the superfields (\ref{latticesuperfields}) and does not involve superspace derivatives of the fields, is:
\beq
\label{Dsuperspace}
S_D = - a^2 \sum_{\mbf} \int d \bar\theta^+ d \theta^-  
K_{I \bar{J}} (U_\mbf, {\bar{U}}_\mbf)\;  
{\bar\Xi}_\mbf^{\bar{J}} \;  \Xi_\mbf^I~,
\eeq
where we generalized  to the multifield case and absorbed $1/a$ factors in the definition of $\Delta_{z, {\bar{z}}}$ appearing in $\Xi, \bar\Xi$.

We note that the axial, fermion number, and involution invariance also allow {\it nonlocal} supersymmetric terms without $\theta$-derivatives of the following form:
\beq
\label{nonlocaltwo}
\int  d \bar\theta^+ d\theta^- \left(   \Xi_\mbf \; \Xi_\nbf 
- {\bar\Xi}_\mbf {\bar\Xi}_\nbf \right) f(U, \bar{U}) ~,
\eeq
for $f(U, \bar{U})$ is an arbitrary function of $U, \bar{U}$, which is odd under the involution. The term (\ref{nonlocaltwo})   
vanishes if $\mbf = \nbf$ due to the fermionic nature of $\Xi$.
Nonlocality, combined with dimensional analysis  implies that  terms of the form (\ref{nonlocaltwo}) are irrelevant in the continuum limit.

Thus, we conclude that the most general local action invariant under $U(1)_V$, $U(1)_A$, $Z_{2 F}$, $I$, and $Z_4$, and not including superspace derivatives, is given by (\ref{Dsuperspace}). The  component expression is immediately seen, upon expansion in superspace, to be: 
\beq
\label{latticeDtermALL}
 S_D = - a^2 \sum_{\mbf}     
& & - K_{I \bar{J}} \; \Delta_z \phi_\mbf^I \cdot \Delta_{\bar z} \phib_\mbf^{\bar{J}}  
+ K_{I \bar{J}} \; F_\mbf^I   {\bar{F}}_\mbf^{\bar{J}}   \nonumber \\
&+& i K_{I \bar{J}} \;  \psib_{-,\mbf}^{\bar{J}} \left[ \Delta_z \psi_{-,\mbf}^I 
+ K^{I \bar{Q}} K_{ML\bar{Q}} \Delta_z \phi_\mbf^M  \cdot  \psi_{-,\mbf}^L \right] \nonumber \\
&-&  i K_{I \bar{J}}  \; \psi_{+,\mbf}^{I} \left[ \Delta_{\bar z} \psib_{+,\mbf}^{\bar{J}}
+ K^{\bar{J} L} K_{L \bar{M} \bar{Q}} \Delta_{\bar z} \phib_\mbf^{\bar{M}}
\cdot  \psib_{+,\mbf}^{\bar{Q}} \right] \nonumber \\
&+& K_{I \bar{J} \bar{L}} F_\mbf^I \psib_{+,\mbf}^{\bar{L}} \psib_{-,\mbf}^{\bar{J}} 
+ K_{\bar{I} J L} {\bar{F}}_\mbf^{\bar{I}} \psi_{-,\mbf}^J \psi_{+,\mbf}^L \nonumber\\
&+& K_{I L \bar{J} \bar{M}} \psib_{+,\mbf}^{\bar{M}} \psi_{-,\mbf}^L \psi_{+,\mbf}^I 
\psib_{-,\mbf}^{\bar{J}}       ~. 
\eeq
For simplicity of notation we denote by $K_{I\bar{J}K}$ the derivative of $K_{I \bar{J}}$ with respect to $U^K$, and similar for higher derivatives, while 
$K^{I \bar{J}}$ is the matrix inverse to $K_{I \bar{J}}$, in the usual continuum notation. The fermion  kinetic terms of $\psi_-, \psib_-$ can be written in terms of the K\" ahler connection $\Gamma^I_{JK} = 
K^{I {\bar{M}}} K_{J \bar{M}, K}$ (and its complex conjugate for $\psi_+, \psib_+$).  All quantities, $K_{I\bar{J}}$, etc., are functions of the scalar fields $\phi_\mbf, \bar\phi_\mbf$  as  indicated in (\ref{Dsuperspace}).

The naive continuum limit of (\ref{latticeDtermALL}) is the   $(2,2)$ nonlinear sigma model action, invariant under the $U(1)_E$, $U(1)_V$,  $U(1)_A$, and involution transformations. 
However, at this point, it should be  clear that eqn.~(\ref{Dsuperspace}), and the corresponding 
finite-dimensional integral over the components of the lattice superfields, with the $Q_-$, $\bar{Q}_+$-invariant measure:
\beq
\label{measure22}
\prod_{\mbf} \prod_{I} d F^I d \bar{F}^{\bar{I}} d \phi^I  d \phib^{\bar{I}} d \psi_-^I d \psib_-^{\bar{I}} d\psi_+^I d \psib_+^{\bar{I}}~,
\eeq
does not define the desired continuum theory. To see this, note that 
we have constructed a fully regulated and supersymmetric lattice  
version of the sigma model, where the continuum global symmetries $U(1)_A$ 
and $U(1)_V$  are both manifest on the lattice. On the other hand, non-Ricci-flat 
continuum sigma models exhibit an anomaly of the axial symmetry, proportional to 
the  first Chern class of the K\" ahler manifold. A regulated version which preserves 
the $U(1)_A$ can not account for the anomaly. There is no puzzle here, of course: a 
study of the perturbative spectrum of (\ref{latticeDtermALL}), see appendix D, reveals 
the presence of doublers, which lead to cancellation of the anomaly. 
This is of
course a direct consequence of the Nielsen-Ninomiya theorem \cite{Nielsen:1980rz}.
To address these 
issues, in the following sections (see also  appendix C and D) we consider in more detail 
the fermion kinetic terms in eqn.~(\ref{latticeDtermALL}), their symmetry, and the definition 
of the lattice partition function of the theory. We note that a version of the lattice action 
(\ref{latticeDtermALL}) for nonlinear sigma models was written before, using a different 
formalism \cite{Catterall:2003uf}.

Let us now summarize  the results of this section. Using our lattice superfield formalism, we constructed a lattice action of the  chiral superfield 2d $(2,2)$ sigma model, which exactly preserves two of the continuum supercharges. The action (\ref{latticeDtermALL}) is the most general local  action consistent with the imposed symmetries, to leading order in the derivative expansion, and for any choice of lattice derivatives.  Upon inspection, it is easy to see that the naive continuum limit of this action coincides with the continuum $(2,2)$ nonlinear sigma model action.  An analysis of the perturbative spectrum of (\ref{latticeDtermALL}) reveals the presence of fermionic (and bosonic, because of the lattice supersymmetry) doublers in the spectrum. We thus have to study various ways to lift the doublers and  discuss  issues related to the definition of the partition function and the quantum continuum limit of the theory.\footnote{The interaction terms in the continuum limit theory with the doublers included as separate ``flavors" most likely violates Lorentz symmetry, as in \cite{Banks:1982ut}; we note, however,  that the arguments of that paper are not directly applicable as the supersymmetry transformations are not  the ones assumed there.}

One possibility\footnote{We thank S. Catterall for making us  think of twisted mass deformations.}  is to use ``twisted" nonlocal mass (i.e., Wilson) terms to lift the doublers. In appendix C, we describe this construction using   our formalism. While this deformation allows the doublers to be lifted and also breaks the anomalous $U(1)_A$, it explicitly violates the exact lattice supersymmetry. 
Since a generic $(2,2)$ sigma model is a renormalizable theory (rather than a superrenormalizable one) with logarithmic divergences arising at every order in perturbation theory, it is not immediately obvious whether the supersymmetric continuum limit can be achieved without fine tuning within this framework; this question deserves further study.

The second mechanism  to lift the doublers uses superpotential nonlocal mass terms.  
These terms  preserve an exact lattice supersymmetry, but explicitly violate some of the global symmetries.  Since some amount of exact supersymmetry is likely to be  helpful in achieving the supersymmetric continuum limit, we devote the rest of the paper to studying the $F$-term deformations and the approach to the continuum limit.   

\subsection{Superpotential, $F$-type Wilson terms, and the continuum limit}
\label{lift2}
As just mentioned, the way to lift the doublers while preserving some exact supersymmetry on the lattice is to incorporate a Wilson mass term via a superpotential. 
A $Q_-$ and $\bar{Q}_+$ supersymmetric $F$-term  is not among the $U(1)_A$, $U(1)_V$, $Z_4$ and $I$ invariants we've already listed. 
An  $F$-term superpotential interaction, including a doubler-lifting nonlocal Wilson mass term,  can be written only if we allow the supersymmetry to be broken  to a linear combination of $Q_-$ and $\bar{Q}_+$. We note that this is accord with our general criterion 2.) from the introduction and with our  discussion of the $(2,0)$ theory of section \ref{20theory}: a superpotential interaction is  an integral over (anti) chiral superspace. The chiral integral's variation under the antichiral supercharge is a total derivative and vice versa. This was the reason that in the $(2,0)$ model an invariant superpotential interaction was not possible. Here, 	however,  we have the opportunity to preserve a linear combination of the chiral and antichiral $Q_-$ and $\bar{Q}_+$ (in other words we let the total derivative variation of the superpotential under $\bar{Q}_+$ cancel that of the anti-superpotential under $Q_-$).

Thus, here we generalize our construction to lattice models where the only supersymmetry is the one generated by $Q_A =Q_- +  \bar{Q}_+  $, i.e. by $\hat{\delta}$ of eqn.~(\ref{22operator}) with $\epsilon^- = - \bar\epsilon^+$ (and $\epsilon^+ = \bar\epsilon^- = 0$). The explicit breaking to the diagonal supersymmetry can be easily  incorporated by introducing a spurion superfield $\Xi_0, \bar\Xi_0$ with the following  $\theta$-dependent ``vev:" 
\beq\label{spurion}
\bar\Xi_0 + \Xi_0 \equiv \bar\theta^+ + \theta^-
\eeq
into an integral over the full superspace.  (We note in passing that a more general
combination is allowed:  $Q_A^\delta = e^{i \delta} Q_- + e^{-i \delta} \bar{Q}_+$.
However, it can be shown that up to a rescaling of parameters of the action by an
overall phase, this is of no consequence to the continuum limit that is obtained.
We therefore set $\delta=0$ in all of our considerations below.)

Consider first the following nonlocal interaction term, invariant under $I$, $Z_{2F}$, and $U(1)_A$, but violating $U(1)_V$ and $Z_4$: 
\beq
\label{Fsuperspace1}
F_{\mbf, \nbf} =  \int d \theta^-  d \bar\theta^+\; (\bar\theta^+ + \theta^-) 
\left( W^\prime(U_\mbf) \; \Xi_\nbf + \bar{W}^\prime ({\bar{U}}_\mbf) \; {\bar\Xi}_\nbf \right)
= (F_1 + F_2)_{\mbf, \nbf},
\eeq  
where prime denotes the derivative of the function $W$ w.r.t. its argument.
The term (\ref{Fsuperspace1}), as indicated above, can be split into two parts, according to their $Z_4$ properties, where:
\beq
\label{Fsuperspace2}
F_{1,\mbf,\nbf} = \int d \theta^- W^\prime (U_\mbf) \; 
\Xi_{\nbf} {}_{\big\vert_{\bar\theta^+ = 0}} - \int d\bar\theta^+ 
\bar{W}^\prime ({\bar{U}}_\mbf) \; {\bar\Xi}_\nbf {}_{\big\vert_{\theta^- = 0}}~,
\eeq
is $Z_4$, as well as $I$, $U(1)_A$, and $Z_{2F}$  invariant. If we recall now that 
$\Xi_{ \big\vert_{{\bar\theta^+}  =0} } =  \psi_+ + \theta^- F$ and the $I$ conjugate relation for $\bar\Xi$ implied by (\ref{latticesuperfields}), we easily find that (\ref{Fsuperspace2}) contains precisely the interactions that form the usual $F$-terms in the continuum limit, with $W(U)$---the superpotential. 

The second part of $F$, called $F_2$ in (\ref{Fsuperspace1}), also respects $I$, $U(1)_A$ and $Z_{2F}$, but violates the discrete $Z_4$ subgroup of Euclidean rotations: 
\beq
\label{Fsuperspace3}
F_{2,\mbf, \nbf} &=& - \int d \bar\theta^+ W^\prime(U_\mbf) \; 
\Xi_{\nbf} {}_{\big\vert_{\theta^- = 0}} + \int d \theta^- 
\bar{W}^\prime ({\bar{U}}_\mbf) \; {\bar\Xi}_\nbf {}_{\big\vert_{\bar\theta^+ = 0}}~\nonumber \\
&=& ~~ i W^\prime(U_\mbf) \; \Delta_{\bar{z}} U_\nbf {}_{\big\vert_{\theta^- = 0}} 
- i {\bar{W}}^\prime({\bar{U}}_\mbf) \;  \Delta_{z} {\bar{U}}_\nbf {}_{\big\vert_{\bar\theta^+ = 0}}  \\
&=& ~~ i W^\prime (\phi_\mbf) \; \Delta_{\bar{z}} \phi_\nbf - i 
\bar{W}^\prime(\phib_\mbf) \; \Delta_{z} \phib_\nbf~.\nonumber 
 \eeq
The lessons from eqns.~(\ref{Fsuperspace1})--(\ref{Fsuperspace3}) 
are that: 

{\it i.)} It is possible to construct interactions that give rise to an $F$ term in the continuum limit and 
preserve an exact lattice supersymmetry.

{\it ii.)} The exact lattice supersymmetry of the $F$ terms requires that these interactions include the  $Z_4$ violating term (\ref{Fsuperspace3}). 
However, also from (\ref{Fsuperspace3}), we see that, just like in quantum mechanics, if we consider the local action, 
$\sum_{\mbf} F_{\mbf, \mbf}$,   the    $Z_4$ violating term $F_2$,    is reduced to a total derivative in the continuum limit. As in quantum mechanics, we expect it  to be irrelevant (if the ultraviolet behavior of the lattice theory is sufficiently soft).

To lift the doublers, we  incorporate a  nonlocal mass terms via the $F$ terms. Thus, the lattice superpotential interaction that we will consider is:
\beq
\label{Fsuperspace}
S_F &=& a^2  \sum_{\mbf} \int d \theta^-  d \bar\theta^+ (\bar\Xi_0 + \Xi_0)
\left( r a \; U_\mbf   \; \Delta^2 \; \Xi_\mbf +r^* a \;  {\bar{U}}_\mbf \;  \Delta^2 \; 
{\bar\Xi}_\mbf \right. \nonumber \\
&& ~ ~ + \left. \sum_{k \ge 2} g_k \; U^{k - 1}_\mbf \; \Xi_\mbf + g^*_k\;  
\bar{U}^{k - 1}_\mbf \; {\bar\Xi}_\mbf  \right)~.
\eeq
Here $r, r^*$ are the complex  Wilson term coefficients, while $g_k, k\ge 2$ are the complex superpotential couplings of unit mass dimension. The local part of the superpotential, implicit in eqn.~(\ref{Fsuperspace}), is:
\beq
\label{wU}
W(U) = \sum_{k \ge 2}\;  {g_k \over k} \: U^k~.
\eeq
The laplacian $\Delta^2$, which appears in the Wilson term, is given explicitly in \myref{llap}.
The full component expression of the $F$-term lagrangian is easily found, 
generalizing to the multifield case, to be:
\beq
\label{latticeFtermlocal}
  S_F &=&a^2 \sum_{\mbf}\left\{  W_{IJ} \; \psi_{-,\mbf}^I  \psi_{+,\mbf}^J + W_I F_\mbf^I 
+ i W_I \Delta_{\bar{z}} \phi_\mbf^I + 
\right. \nonumber \\
&~& \left.+ \bar{W}_{\bar{I} \bar{J}} \; \psib_{+,\mbf}^{\bar{I}} 
\psib_{-,\mbf}^{\bar{J}}+ \bar{W}_{\bar{I}}  
\bar{F}_\mbf^{\bar{I}} - i \bar{W}_{\bar{I}} \Delta_z \phib_\mbf^{\bar{I}} \right. \\
&~&\left. + r  a \;  \psi_{-,\mbf}^I \Delta^2 \psi_{+,\mbf}^I + r a \; 
\phi_\mbf^I \Delta^2 F_\mbf^I  \right. \nonumber \\
&~& \left.+ r^* a\; \psib_{+,\mbf}^{\bar{I}} \Delta^2 \psib_{-,\mbf}^{\bar{I}}
+ r^*  a \; \phib_\mbf^{\bar{I}} \Delta^2 \bar{F}_\mbf^{\bar{I}} \right\} \nonumber~.
\eeq
(In appendix \ref{graphs}, we demonstrate how to incorporate the
nonlocal Wilson terms into the superpotential by an appropriate
modification of \myref{wU}.)
We note that the three-derivative bilinear scalar terms, of the form 
$i \phi_\mbf \Delta^2 \Delta_{\bar{z}} \phi_\mbf$, which would seem to appear in the Wilson term from (\ref{Fsuperspace}) vanish after summation over the periodic lattice. As usual, $W_{I}, W_{I J}$ denote derivatives of the superpotential (\ref{wU}) with respect to the
fields $U^I$, evaluated at $\phi_\mbf$. The naive continuum limit of the lattice action is, by inspection of (\ref{latticeFtermlocal}) the usual superpotential interaction in the $(2,2)$ continuum theory. The irrelevant    nonlocal mass terms, as we show in appendix D, lift the doublers by giving them mass of order $|r| a^{-1}$ and ensure that the spectrum of the lattice theory matches that of the continuum.

  It is important to note that the exact lattice supersymmetry and the other global symmetries do not require the action to have the form (\ref{latticeFtermlocal}) (with $D$-term  (\ref{Dsuperspace})). In fact, if  we were to ask for the most general  involution, fermion number,  $U(1)_A$, and $Q_A$ invariant action of lowest dimension (i.e. local and without superspace derivatives on the fields), 
we would find the following relevant local term:
\beq
\label{Gsuperspace1}
G_\mbf =  \int d\theta^- d \bar\theta^+ (\bar\theta^+  + \theta^-) \left( {\cal{G}}(U_\mbf, {\bar{U}}_\mbf)\;  \Xi_\mbf+ {\cal{G}}^*({\bar{U}}_\mbf, U_\mbf) \; \bar\Xi_\mbf \right) ~, 
\eeq
where ${\cal{G}}$ is now an arbitrary function of $U$ {\it and} $\bar{U}$---as opposed to $W^\prime(U)$ of (\ref{Fsuperspace1})---of unit mass dimension. An additional marginal term would have  form similar to  (\ref{Gsuperspace1}), but ${\cal{G}}$ would be then linear in derivatives.
Consider now the structure  of (\ref{Gsuperspace1}) in more detail.  As in the case of (\ref{Fsuperspace1}), we can decompose $G = G_1 + G_2$, where $G_1$ preserves the discrete  rotation $Z_4$ symmetry, while $G_2$ does not:
\beq
\label{Gsuperspace2}
G_1 &=& \int d \theta^- {\cal{G}} (U, \bar{U}) \Xi_{\big\vert_{\bar\theta^+ = 0}} + \;  {\rm I. c.} \;= \; {\cal{G}}(\phi, \bar{\phi}) \; \psi_- \psi_+ + {\cal{G}}_{, \phi} (\phi, \phib) \; F + \;  {\rm I. c.}  \\
G_2 &=& - \int d \bar\theta^+ {\cal{G}} (U, \bar{U}) \Xi_{\big\vert_{\theta^- = 0}} + \; {\rm I.c.} \; =\;  i {\cal{G}}(\phi, \phib)\; \Delta_{\bar{z}} \phi + {\cal{G}}_{, \phib} (\phi, \phib)\; \psib_+ \psi_++ \; {\rm I.c.} \nonumber
\eeq
Thus, $G_1$ contains nonholomorphic corrections to the superpotential, while  $G_2$ consists of $Z_4$ violating terms, related to the nonholomorphic $F$ terms in $G_1$ by the exact lattice supersymmetry.  Thus, the symmetries of the lattice action are not sufficient to guarantee that the $(2,2)$ supersymmetric continuum limit is achieved. Moreover, there are no additional symmetries,  within the present approach,  that can forbid terms like (\ref{Gsuperspace1}) but allow (\ref{Fsuperspace}). We conclude that the action with the desired $(2,2)$ supersymmetric continuum limit is not generic.

We thus need to address the following question: given a supersymmetric lattice theory with  action given by (\ref{Dsuperspace}), and with superpotential (\ref{Fsuperspace}), are nonholomorphic, non-Lorentz-invariant  terms in the action of the form (\ref{Gsuperspace2}) generated  quantum mechanically?

In this paper, we will only address the case of Landau-Ginzburg (or Wess-Zumino) models. In other words, we consider supersymmetric lattice theories with flat K\" ahler metric, $K_{I \bar{J}} = \delta_{I \bar{J}}$, and some polynomial superpotential. We hope to address the (significantly more involved) renormalization of models with non-trivial K\" ahler metric, both with supersymmetric and non-supersymmetric Wilson terms in the future.

To study the quantum continuum limit, we consider in some detail 
the counting of divergences in lattice perturbation theory in the Landau-Ginzburg model,
in appendix D. As discussed there, the lattice power counting rules 
show that the lattice introduces several extra divergent graphs 
(compared to the continuum case) due to the higher-derivative vertices 
induced by the supersymmetrization of the Wilson term and by the 
interaction term (\ref{surface}).   However, as   
also shown in appendix D, the exact lattice supersymmetry is 
sufficient to ensure that all divergent as $a \rightarrow 0$   
lattice graphs cancel and that  the lattice theory is finite.  
That is, the net sum of all lattice perturbation theory Feynman diagrams
contributing to any proper vertex
can be seen to have a negative  degree of divergence 
(determined by lattice power-counting),
also due to the exact lattice supersymmetry. 
This ensures that  the lattice and continuum perturbation expansions are identical in the $a \rightarrow 0$ 
limit\footnote{For general theorems on asymptotic expansions of 
lattice Feynman integrals and  lattice power counting rules, see  \cite{Reisz:1987da}.} and 
that the continuum $(2,2)$ supersymmetric continuum 
limit is achieved without  any fine tuning.  We note that the finiteness due to 
the exact lattice supersymmetry was crucial to the argument.

Finally, we note that the $Z_4$ violating term  in the lattice action (\ref{latticeFtermlocal}), despite appearing relevant as written, is, in fact, an irrelevant operator. This can be seen, again, by using periodicity of the lattice. In  the $W(\phi) = g \phi^3/3$ case (quadratic terms in the superpotential do not contribute to the $Z_4$ violating term due to periodicity on the lattice),   the $Z_4$-violating term in the action can be identically written as:
\beq
\label{surface}
a^2 \sum_{\mbf}  \; g \phi_\mbf^2 \Delta_{\bar{z}} \phi_\mbf \equiv {1\over 6} \; a^2 \sum_{\mbf} \;  \;a^2 g \left( (\Delta^+_1   \phi_\mbf)^3- i (\Delta^+_2   \phi_\mbf)^3\right),
\eeq
with $\Delta^+_i$---the forward derivative in the $i$-th direction and we used symmetric differences in $\Delta_{\bar{z}}$. It is easy to check that the  $Z_4$ violating term in the lattice action is of order $a^2$  for any power $k > 2$ in the superpotential, but the identity is not as simple as for the $k=3$ case above. The irrelevant Lorentz violating term has no effect on the continuum theory due to the 
finiteness of the lattice theory (another way to see that the nonholomorphic terms of (\ref{Gsuperspace2}) are irrelevant, as they are related to the Lorentz violating ones by the lattice supersymmetry).  

Thus, as mentioned in section \ref{outs}, the results of
this section demonstrate that the $(2,2)$ LG models can be simulated on the lattice.

\vspace{20pt}

\noindent
{\bf \large Acknowledgements}

\vspace{5pt}
We would like to thank Kentaro Hori and Simon Catterall for
useful discussions. This work was supported by the National Science and Engineering 
Research Council of Canada and the Ontario 
Premier's Research Excellence Award.

\noindent

\appendix

\vspace{20pt}

\noindent
{\bf \large Appendices}

\vspace{5pt}

\section{The transfer matrix and Witten index of the lattice supersymmetric quantum mechanics}
Consider the lattice partition function (\ref{QMZ}) after a convenient change of
 fermionic variables $\psi^i= \bar\eta^{i+1}$, $\chi^i = \eta^i$: 
\beq
\label{QMZappx}
Z_N =  c^N \prod\limits_{i=1}^N \int d\bar\eta^i  d \eta^i \int\limits_{- \infty}^{\infty} d x^i \;  e^{- S}~, 
\eeq
with the action (\ref{QMaction2}) in terms of the new variables:   
\beq
\label{QMaction2appx}
S = \sum\limits_{i=1}^N \left[   {a\over 2}  \left(  {x^{i+1} -  x^{i} \over a} + h^\prime (x^{i+1}) \right)^2  -   \bar\eta^{i+1}  \eta^i 
 \left( a h^{\prime \prime} (x^{i}) + 1  \right) + \bar\eta^i   \eta^i \right]~
\eeq
The lattice  action above  is not reflection positive. Nevertheless, we will show that in the continuum limit it defines a hermitean hamiltonian acting on a positive norm Hilbert space---the hamiltonian of supersymmetric quantum mechanics.
The lack of reflection positivity comes from the cross term in the expansion of the square in the bosonic action (which would be a total derivative, ${\partial\over \partial t} h$ in the continuum limit). The presence of this term is required by the exact lattice supersymmetry of the action. It is natural to expect that in the continuum limit its effect will be irrelevant; this is what we want to demonstrate here.\footnote{As already mentioned in section \ref{susyqm}, the choice of discretization of the  derivative in (\ref{QMaction2appx}) corresponds to   a Wilson term with $r \equiv 1$. This is the choice where only nearest-neighbor interactions occur, for which the construction of the transfer matrix and Hamiltonian from the Euclidean lattice action is most straightforward.}

To construct the transfer matrix and hamiltonian, see \cite{Montvay}, we first introduce, at each time slice, a Hilbert space which is  a tensor product of a bosonic and fermionic space. The bosonic Hilbert space is that of square integrable functions on the line. We use the  basis of position eigenstates, $\{\vert x \rangle, \; \langle x^\prime | x \rangle = \delta(x^\prime - x) \}$,   where the momentum and position operators, $\[ \hat{p}, \hat{q} \] = - i$, act as $\hat{q} \vert x \rangle = \vert x \rangle x$ and $e^{i \hat{p} \Delta } \vert x \rangle= \vert x + \Delta \rangle$ 
(note that we continue using the dimensions of section \ref{susyqm}: $x$ has mass dimension $-1/2$, $a$ has dimension of length, while the superpotential $h(x)$ is dimensionless). The fermionic Hilbert space is two dimensional and is  spanned by the vectors $\vert 0 \rangle$ and $\vert 1 \rangle$. The  fermionic creation and annihilation operators obey $\{\hat{b}^\dagger, \hat{b} \}= 1$, such that $\hat{b} \vert 0 \rangle = 0$, $\vert 1 \rangle = \hat{b}^\dagger \vert 0 \rangle$. 
The fermionic coherent states are defined as $\vert \eta \rangle\equiv \vert 0 \rangle + \vert 1 \rangle  \eta$, $\langle \eta \vert = \langle 0 \vert + \bar\eta \langle 1 \vert$, where $\eta$ and $\bar\eta$ are Grassmann variables. We then recall the usual relations for the decomposition of unity, $\langle \eta^\prime \vert \eta \rangle = e^{\bar\eta^\prime \eta}$; ${\hat{1}} = \int  d\bar\eta d \eta e^{- \bar\eta\eta} \vert \eta \rangle \langle \eta \vert$, and  for 
traces\footnote{We note in passing that  the first of the two 
relations that follow can, along with the expression for the 
$T$-matrix (\ref {Tmatrix3}), be used to give a path integral 
representation of the finite temperature partition function as 
well. Clearly,  this amounts to switching the sign of $\eta$ only 
in the second term of eqn.~(\ref{QMaction2appx}) (equivalently, 
the second term of (\ref{Tmatrix2})) at a single lattice site only (despite appearances, eqn.~(\ref{Tmatrix}) shows that discrete time translation invariance is not broken, since the point can be freely moved around). The exact supersymmetry is then globally broken by the boundary conditions, but locally present.}  of operators ${\cal{O}}$ on the fermionic Hilbert space: 
Tr$\; {\cal{O}} = \int  d\bar\eta d \eta e^{- \bar\eta\eta} \langle \eta\vert {\cal{O}} \vert - \eta \rangle$;   Tr$(-1)^F {\cal{O}} = \int  d\bar\eta d \eta e^{- \bar\eta\eta} \langle \eta\vert {\cal{O}} \vert  \eta \rangle$, with  $(-1)^F \vert 0 \rangle = \vert 0 \rangle$.
We then define the transfer matrix  by the equality:
\beq
\label{Tmatrix}
Z_N &=& c^N \prod\limits_{i=1}^N \int d\bar\eta^i  d \eta^i \int\limits_{- \infty}^{\infty} d x^i \;  e^{- S} \equiv {\rm Tr} \; (-1)^F \; \hat{T}^N   = \\
&=&  \prod\limits_{i=1}^N \int d\bar\eta^i  d \eta^i e^{- \bar\eta^i \eta^i} \int\limits_{- \infty}^{\infty} d x^i \times \nonumber \\
&\times& \langle \eta^N, x^N \vert \hat{T} \vert \eta^{N-1}, x^{N-1} \rangle \langle \eta^{N-1}, x^{N-1} \vert \hat{T} \vert \eta^{N-2}, 
x^{N-2} \rangle \times \ldots \nnn
&& \times \langle \eta^2, x^2 \vert \hat{T} \vert \eta^{1}, 
x^{1} \rangle \langle \eta^1, x^1 \vert \hat{T} \vert \eta^{N}, x^{N} \rangle \nonumber~,
\eeq
or, equivalently, through its matrix elements:
\beq
\label{Tmatrix2}
 \langle \eta^{i+1}, x^{i+1} \vert \hat{T} \vert \eta^{i}, x^{i} \rangle &=& \\
&c \;  \exp&\left[-  {a\over 2}  \left(  {x^{i+1} -  x^{i} \over a} + h^\prime (x^{i+1}) \right)^2  +  \bar\eta^{i+1}  \eta^i 
 \left( 1+ a h^{\prime \prime} (x^{i})   \right)  \right]\nonumber ~.
\eeq
Using  $\langle \eta^\prime | 1 - X \hat{b}^\dagger \hat{b} |\eta    \rangle = e^{(1-X) \bar\eta^\prime \eta}$, it is straightforward to check that the 
 $\hat{T}$ operator with matrix elements (\ref{Tmatrix2}) is given by:
\beq
\label{Tmatrix3}
\hat{T} \; = \; c \int\limits_{-\infty}^\infty d z\; \exp\left( - {a \over 2} \left( {z \over a} + h^\prime (\hat{q})  \right)^2 \right)  \; \exp \left(i z \hat{p} \right)\;  \left(1 + a h^{\prime\prime}(\hat{q})\; \hat{b}^\dagger \hat{b} \right)~.
\eeq
In the small-$a$ limit, this operator becomes, with $c = (2 \pi a)^{-{1\over 2}}$:
\beq
\label{Tmatrix4}
\hat{T} &=& \exp\left( - {a \over 2} \hat{p}^2 - {a \over 2} i \left( h^\prime(\hat{q}) \;  \hat{p} + \hat{p} \; h^\prime(\hat{q}) \right) + a h^{\prime\prime}(\hat{q})\; \left( \hat{b}^\dagger \hat{b} - {1 \over 2} \right) \right) e^{  {\cal{O}}(a^2)}\nonumber \\
&=& e^{ h(\hat{q})} \; \exp\left( - a \left( {\hat{p}^2 \over 2}  + {h^\prime(\hat{q})^2\over 2}  -    {h^{\prime\prime} (\hat{q}) \over 2} \[ \hat{b}^\dagger , \hat{b} \] \right) \right) \; e^{- h(\hat{q})}  \;e^{  {\cal{O}}(a^2)}~.
\eeq
The term in the middle exponent, proportional to a single power of the lattice spacing, is easily recognized as the Hamiltonian of the supersymmetric quantum mechanics, $H_{SQM}$. The limit of small lattice spacing should, of course, be understood in the weak sense (as for arbitrarily small $a$ there always exist large enough $x$ such that the order $a^2$ term is important; however, for a  potential sufficiently strong at infinity these values of $x$ make an exponentially small contribution to the path integral). As eqn.~(\ref{Tmatrix4}) shows, in the $a \rightarrow 0$ limit, the $\hat{T}$ matrix (\ref{Tmatrix3}), with matrix elements (\ref{Tmatrix2}), is  conjugate to ${\hat{T}}_{cont.} \equiv e^{- a H_{SQM}}$: 
\beq
\label{Tmatrix5}
\hat{T} \simeq e^{  h(\hat{q}) }  \;  e^{ - a H_{SQM} } \; e^{ -h( \hat{q} ) } ~.
\eeq
Inserting (\ref{Tmatrix5})  into (\ref{Tmatrix}), one observes that the $e^{\pm h}$ factors cancel out of the partition function, and we are left with the usual ``naive" path integral representation for  the partition function. 

We thus conclude that, in the continuum limit the two discretizations---the ``naive"  and the supersymmetric, with $\hat{T}$ of eqn.~(\ref{Tmatrix2}), are equivalent: both converge  to  
$ Z =  {\rm Tr} \; (-1)^F e^{ - \beta H_{SQM}}$. 
The supersymmetric discretization, however, enjoys nice properties already at finite $N$ and $a$; for example, as we show below, it gives the correct value of the Witten index already at finite lattice spacing. Moreover, if  one was interested in numerical computations of correlation functions in  supersymmetric quantum mechanics, one would find much faster convergence to the supersymmetric continuum limit in the case of a supersymmetric discretization. 

We now elaborate on the property of the supersymmetric lattice action alluded to above, and show  how the exact lattice supersymmetry of  (\ref{QMZ}) (or of (\ref{QMZappx})) implies that  the correct value for the Witten index is obtained already  at finite lattice spacing.
To this end we take (\ref{QMZ}) in the form:
\beq
\label{WI1}
Z_N = (2 \pi a)^{-{N\over 2}} \int \prod\limits_{i = 1}^N d x^i d\chi^i d \psi^i \exp \left( - {a \over 2} \; ~
{\cal{N}}^i {\cal{N}}^i(x) - a  \; \chi^i {\partial {\cal{N}}^i (x) \over \partial x^j} \psi^j \right)
\eeq 
where ${\cal{N}}^i(x) \equiv  h^\prime(x^i) + (x^i - x^{i-1})/a$ is the Nicolai variable 
(\ref{Nicolai1}) (summation over repeated indices is understood in this discussion). Both the measure and action are invariant under the nilpotent supersymmetry generated by $Q$, which acts as implied by (\ref{latticesuperfieldsQ}): $\delta_\epsilon x^i = \epsilon \psi^i$, $\delta_\epsilon \psi^i = 0$, $\delta_\epsilon \chi^i = - \epsilon {\cal{N}}^i$. $Q$-invariance of the measure and action imply that  (schematically) 
$ \int d x d \chi d \psi \;e^{-S}\;  \delta_\epsilon X(x,\chi,\psi)  = 0$, i.e., that  correlation functions of $Q$-exact operators vanish.

Consider now the $Q$-variation of  a particular  $X(x,\chi,\psi)$, chosen as
$X = -  \chi^i g^i(x)$, where $g^i$ is some function of the $x^k$:  $\delta_\epsilon X = \epsilon ({\cal{N}}^i g^i + \chi^i { \partial g^i \over \partial x_j} \psi^j)$.  The crucial point is that this $Q$-variation of $X$ is the same as the variation of the action, $S = {1\over 2}   ( {\cal{N}}^i(x))^2 +  \chi^i \; {\partial {\cal{N}}^i (x) \over \partial x^j } \; \psi^j $, under an $x$-dependent shift of the Nicolai variable:  $\delta_g {\cal{N}}^i = g^i(x)$: $\epsilon \delta_g S =  \delta_\epsilon X$. Therefore, under a shift of the Nicolai variable, the change of $Z_N$ is a correlation function of a $Q$-exact operator. Since such correlators vanish, the  conclusion of the previous paragraph implies that  $Z_N$ is invariant under deformations of the Nicolai variable 
(provided they do not change the asymptotics of $h$ at infinity). In particular, upon choosing $g^i = - (x^i - x^{i-1})/a$,   the  lattice sites decouple and we obtain a simple expression for $Z_N$:
\beq
\label{WI2}
Z_N &=& \left(  \int\limits_{-\infty}^{+\infty} {dx\over \sqrt{2 \pi a}} \int d \chi d \psi \; \exp\left( - {a \over 2}\; h^\prime(x)^2 - a h^{\prime \prime} (x) \;  \chi \psi \right) \right)^N  \\
&=&   \left(  \int\limits_{-\infty}^{+\infty}  {dx\over \sqrt{2 \pi}}  \; h^{\prime \prime} (x) \;  \exp\left( - {1 \over 2} \; h^\prime(x)^2 \right)  \right)^N \; =\;  \left(  \sum_{x^*} {h^{\prime\prime}(x^*) \over |h^{\prime\prime}(x^*)|} \right)^N . \nonumber
\eeq
where the sum over the critical points $x^*$ of the superpotential $h(x)$, $\sum_{x^*} {h^{\prime\prime}(x^*) \over |h^{\prime\prime}(x^*)|}$,   takes values   $\pm 1$ or $ 0$, depending on whether the superpotential is ``odd" or ``even" at infinity. This sum, in fact,  
equals the Witten index of the continuum   quantum mechanics with superpotential~$h$.
 Thus, the supersymmetric lattice path integral precisely reproduces, already at finite $N$,  the continuum value of the Witten index, for odd total number of lattice points $N$. This is to be expected: recall that in the continuum one calculates  the index by compactifying on a Euclidean circle with periodic boundary conditions. One then calculates the determinant of the quantum fluctuations around field configurations on which the path integral is localized
($\dot{x} = h^\prime(x) = 0$). A complete cancellation  between the nonzero modes occurs only if an odd (of  course, infinite) number of bosonic modes is present (a zero mode and an even number of Kaluza-Klein modes). In our lattice regularization this corresponds to having an odd number of lattice sites.

\section{A-type lattice superfield kinematics}
In the standard superspace notation, the supersymmetry generators in (\ref{22susyalgebra}) are represented as differential operators acting on superfields:
\beq
\label{22generators}
Q_{\pm} &=& {\partial \over \partial \theta^{\pm}} + i \bar\theta^{\pm} \partial_\pm ~,\nonumber \\
\bar{Q}_{\pm} &=& - {\partial \over \partial \bar\theta^{\pm}} - i \theta^{\pm} \partial_\pm~. 
\eeq
A supersymmetry transformation with parameters $\epsilon^{\pm}$, $\bar\epsilon^{\pm}$ is generated by the action of:
\beq
\label{22operator}
\hat{\delta} = \epsilon^- Q_- + \epsilon^+ Q_+ - \bar\epsilon^- \bar{Q}_- - \bar\epsilon^+ \bar{Q}_+~
\eeq
on superfields.
Supersymmetric  actions  are   written as integrals over superspace of various superfields. The variation of a superspace Lagrangian density is found by  acting with the differential operator  $\delta$. Since spacetime derivatives appear in the $Q$'s, the Lagrangian densities are $\delta$-invariant only  up to total derivative terms. The validity of the Leibnitz rule for spacetime derivatives is thus crucial for preserving supersymmetry; as stated many times above, this is the main obstacle for preserving the supersymmetry algebra on the lattice (in interacting models). 
The appearance of the  ``troublesome" spacetime derivatives in (\ref{22generators}) can, in its  turn, be traced to the non-anticommutativity of the supercharges. 
Now, recall that the supersymmetry generators can also be written  in the following  form:
\beq
\label{22operatorsNEW}
Q_+ &=& e^{- X_+} \;   {\partial \over \partial \theta^+} \; e^{X_+} \equiv {\partial \over \partial \theta^+} + \left( {\partial \over \partial \theta^+} X_+\right),~ ~{\rm where} ~ X_+ \equiv i \theta^+ \bar\theta^+ \partial_+ ~, \nonumber \\
Q_- &=& e^{- X_-} \;   {\partial \over \partial \theta^-} \; e^{ X_-} \equiv {\partial \over \partial \theta^-} + \left( {\partial \over \partial \theta^-} X_-\right),~ ~{\rm where} ~ X_- \equiv i \theta^- \bar\theta^- \partial_- ~, \\
\bar{Q}_+ &=& - e^{X_+} \;   {\partial \over \partial \bar\theta^+} \; e^{-X_+} \equiv - {\partial \over \partial \bar\theta^+} + \left( {\partial \over \partial \bar\theta^+} X_+\right),~  \nonumber \\
\bar{Q}_- &=& - e^{X_-} \;   {\partial \over \partial \bar\theta^-} \; e^{ -X_-} \equiv - {\partial \over \partial \bar\theta^-} + \left( {\partial \over \partial \bar\theta^-} X_-\right).~  \nonumber
\eeq
It follows from the above equation that if two supercharges anticommute, with an appropriate change of coordinates,  the corresponding differential operators can be represented  simply as derivatives with respect to the odd superspace coordinates. For example, it follows from eqn.~(\ref{22operatorsNEW}) that the anticommuting operators $\bar{Q}_+$ and $\bar{Q}_-$ (our  ``$B$-type") can be represented solely by supercoordinate derivatives upon conjugation:
\beq
\label{Bconjugation}
\bar{Q}_\pm = e^{X_+ + X_-} \; \bar{Q}^B_\pm \; e^{- X_+ - X_-} ~, ~~ \bar{Q}^B_{\pm} = - {\partial \over \partial \bar\theta^{\pm}}~.
\eeq
We can also see from (\ref{22operatorsNEW}) that supersymmetry generators whose anticommutator involves  a spatial derivative can not be conjugated to purely supercoordinate derivative---for example, representing $Q_+$ and $\bar{Q}_+$ by pure $\theta$ derivatives requires opposite  conjugations, with $e^{X_+}$ and $e^{-X_+}$, respectively. For further use, we note that the supercovariant derivatives can be written similar to (\ref{22operatorsNEW}):
\beq
\label{supercovariant}
D_+ &=& e^{X_+}\; {\partial \over \partial \theta^+}\; e^{-X_+}, ~~ D_- = e^{X_-} \;{\partial \over \partial \theta^+}\; e^{-X_-} \nonumber \\
\bar{D}_+ &=&  - e^{-X_+} {\partial \over \partial \bar\theta^+} e^{X_+}, ~~ \bar{D}_- = - e^{-X_-} {\partial \over \partial \bar\theta^-} e^{X--}~.
\eeq

Clearly, for the ``$A$-type," a simultaneous representation of the $\bar{Q}_+$ and $Q_-$ as purely supercoordinate derivatives also exists:
\beq
\label{Aconjugation}
Q_- &=& e^{X_+ - X_-} \;  Q_-^A \; e^{X_- - X_+} ~~, ~~ {\rm with} ~ Q_-^A \equiv {\partial \over \partial \theta^-}~, \nonumber \\ 
\bar{Q}_+ &=& e^{X_+ - X_-}\; \bar{Q}_+^A \; e^{X_- - X_+} ~~, ~~ {\rm with}~ \bar{Q}^A_+ \equiv -{\partial \over \partial \bar\theta^+} ~. 
\eeq
The form of the remaining $Q_+$ and $\bar{Q}_-$ after conjugation with $e^{X_- -X_+}$ can easily be worked out and seen to involve spacetime derivatives. Thus, these two remaining supercharges will not be exact symmetries  of  an interacting  $A$-type lattice action.

Our next task is to construct  the type-$A$ lattice superfields.  As we are interested in theories whose continuum limit is a theory of chiral superfields, we begin by transforming 
the familiar continuum chiral superfields into the basis where   $Q^-$ and $\bar{Q}^+$ act by shifts of their respective odd superspace coordinates. 

A chiral superfield $\Phi$ obeys the covariant constraint $\bar{D}_\pm \Phi = 0$. This constraint is easy to solve in the chiral basis, defined by $Q = e^{ - X_- - X_+} \; Q^\chi \; e^{X_- +X_+}$,  where $Q_\pm$ and $\bar{D}_{\pm}$ are 
simultaneously represented in terms of pure $\theta$ derivatives. 
The chiral superfield's component expansion is then easily seen to be:
\beq
\label{chiralcontinuum}
\Phi(x^\pm, \theta^\pm, \bar\theta^\pm) &=& 
e^{- X_+- X_-}\;  \Phi^\chi (x^\pm, \theta^\pm) \; e^{X_+ +X_-}  \nnn
&=&   \phi(y^\pm) + \theta^+ \psi_+ (y^\pm) + \theta^- \psi_-(y^\pm) + \theta^+ \theta^- F(y^\pm) ~.
\eeq
In (\ref{chiralcontinuum}) we denote by $y^\pm$ the coordinate $y^\pm = x^\pm - i \theta^\pm \bar\theta^\pm$. 
To transform to the $A$-type basis   we   conjugate the fields and charges as 
  $\Phi^A =$ $e^{X_--  X_+} \; \Phi(x^\pm, \theta^\pm, \bar\theta^\pm)  \; e^{X_+-X_-}$.
Thus, the $A$-basis  conjugated chiral superfield is: 

{\flushleft$ \Phi^A \; (x^{   +}, x^{   -}, \theta^\pm, \bar\theta^\pm) =$}
\beq
\label{chiralcontinuumconjugated}
 &=&e^{ X_- -   X_+}\; \Phi (x^\pm, \theta^\pm, \bar\theta^\pm) \; e^{X_+ -X_-} \nonumber \\
&=&e^{ X_- -   X_+}\; \left[  \phi(y^\pm) + \theta^+ \psi_+ (y^\pm) + \theta^- \psi_-(y^\pm) + \theta^+ \theta^- F(y^\pm) \right] \; e^{X_+ -X_-}\\
&= &  \phi(x^+ - 2 i \theta^+ \bar\theta^+  ,  x^- ) + \theta^+ \psi_+ (x^+ - 2  i \theta^+ \bar\theta^+ ,  x^- )  \nonumber \\
&~& + \theta^- \psi_-(x^+- 2 i \theta^+ \bar\theta^+,  x^- ) + \theta^+ \theta^- F(x^+- 2  i \theta^+ \bar\theta^+ ,  x^-) \nonumber\\
&=& \phi      - 2 i \theta^+ \bar\theta^+ \partial_+ \phi +  \theta^+ \psi_+ + \theta^- \psi_-   - 2 i \theta^- \theta^+  \bar\theta^+ \partial_+ \psi_- + \theta^+ \theta^- F~,
\nonumber
\eeq
where all component fields in the last line are functions of $x^\pm$. 
In the $A$-representation above, the $Q^-$ and $\bar{Q}^+$ supercharges now act simply by shifting $\theta^-$ and $\bar\theta^+$.
We can rewrite the chiral superfield (\ref{chiralcontinuumconjugated}) as follows:
\beq
\label{chiral2}
\Phi^A  =  \left( \phi + \theta^- \psi_-  \right) + \theta^+ \left( \psi_+ - 2 i \bar\theta^+ \partial_+ \phi + \theta^- F + 2 i  \theta^- \bar\theta^+ \partial_+ \psi_- \right)  \equiv U + \theta^+ \Xi  ~.
\eeq
The two terms in brackets in (\ref{chiral2}) are $\theta^-, \bar\theta^+$-dependent superfields, with irreducible action of the $A$-type supercharges.
Thus, under the action of $Q^-$ and $\bar{Q}^+$ alone, the chiral superfield $\Phi$ splits into two irreducible components---the $U$ and $\Xi_+$ superfields: 
\beq
\label{UandXi}
U(x^\pm, \theta^-) &=& \phi + \theta^- \psi_- ~\nonumber \\
\Xi  (x^\pm, \theta^-, \bar\theta^+) &=& \psi_+ - 2 i \bar\theta^+ \partial_+ \phi + \theta^- F + 2 i  \theta^- \bar\theta^+ \partial_+ \psi_- ~.
\eeq

Similarly, since an antichiral superfield can be written as $\bar\Phi(x^\pm, \theta^\pm, \bar\theta^\pm) = \phib (\bar{y}^\pm) - \bar\theta^- \psib_- (\bar{y}^\pm) -\bar\theta^+ \psib_+  (\bar{y}^\pm) + \bar\theta^- \bar\theta^+ \bar{F}  (\bar{y}^\pm)$, with $\bar{y}^\pm =  x^\pm + i \theta^\pm \bar\theta^\pm$, 
it follows  that an antichiral superfield in the $A$ basis is given by:
\beq
\label{antichiral2}
\bar\Phi^A  &=& e^{X_- -X_+}\; \bar\Phi(x^\pm, \theta^\pm, \bar\theta^\pm)  \;  e^{X_+ -X_-} = \phib(x^+, x^- + 2 i \theta^- \bar\theta^-) + \ldots \\
&=& \left( \phib - \bar\theta^+ \psib_+  \right) - \bar\theta^- \left( \psib_- + 2 i \theta^- \partial_-\phib - \bar\theta^+ \bar{F} - 2 i \theta^- \bar\theta^+ \partial_- \psib_+ \right)  \equiv \bar{U} - \bar\theta^- \bar\Xi  ~. \nonumber
\eeq
We see that the antichiral field $\bar\Phi$ also splits in to two superfields that transform irreducibly under the type-$A$ supercharges.
The superfields  $\bar{U}$ and $\bar\Xi$ are  defined by  eqn.~(\ref{antichiral2}). 
Their properties under I-conjugation were discussed (for their lattice version)
in section \ref{choice} above; there it was seen that this conjugation is not the usual Minkowski space 
complex conjugatation of $U$ and~$\Xi$.

As explained in the introduction, our lattice theory (prior to lifting spectrum doublers) 
will preserve exactly the supersymmetries represented by shifts of   $\theta^-$ and $\bar\theta^+$. 
We will construct $Q^-$ and $\bar{Q}^+$ invariants using  a lattice version of the superfields $U, \bar{U}$, $\Xi, \bar\Xi$. Since the action of these supersymmetries does not involve any spacetime derivatives, the latticization,  including interactions, will not be in conflict with the $Q^-$ and $\bar{Q}^+$ supersymmetries.

We now introduce two pairs of 
  bosonic ($U, \bar{U}$)  and  fermionic ($\Xi, \bar\Xi$) lattice superfields by simply latticizing their continuum definitions (\ref{antichiral2}) and (\ref{UandXi}). 
Before introducing the lattice, we   rotate to Euclidean space, by  
replacing $x^0$ by $- i x^2$. Thus, $\partial_0 = i  \partial_2$. We 
also introduce the complex notation $z = x^1 + i x^2$,  
$\partial_z = (\partial_1 - i \partial_2)/2$. Since $\partial_\pm = (\partial_0 \pm \partial_1)/2$, we find that $\partial_+ = \partial_{\bar{z}}$, $\partial_- = -\partial_{z}$.

We let our lattice have sites labeled by integer valued 
vectors $\mbf$, $m_{1, 2} = 1, \ldots, N$, with $m_i + N \equiv m_i$. 
The continuum derivatives in the $i$-th direction  are replaced by
lattice derivatives. The symmetries  and   conjugation rules  of the
Euclidean formulation  will be discussed shortly; here we only note 
that any choice of derivative is compatible with the 
symmetries (\ref{superfieldsymmetries}). Thus, until we make a choice of 
lattice derivatives, we denote the derivatives in the two 
directions by $\Delta_1$ and $\Delta_2$. We also denote 
$\Delta_z = \Delta_1 - i \Delta_2$ and $\Delta_{\bar{z}} = \Delta_1 + i \Delta_2$ 
(without factors of $1/2$), so that the discrete form of 
$2 \partial_+$ is $\Delta_{\bar z}$ and that of $2 \partial_-$ is 
$- \Delta_z$. In fact, making these replacements in (\ref{UandXi}) and 
their I-conjugates, we arrive at the lattice superfields (also given in eqn.~(\ref{latticesuperfields})):
\beq
\label{latticesuperfieldAppx}
U_\mbf &=& \phi_\mbf+ \theta^- \psi_{-,\mbf}~, \nonumber \\
\Xi_\mbf&=& \nonumber \psi_{+,\mbf} - \bar\theta^+ i \Delta_{\bar z} \phi_\mbf + \theta^- F_\mbf +  i  \theta^- \bar\theta^+  \Delta_{\bar z} \psi_{-,\mbf}~ , \nonumber \\
{\bar{U}}_\mbf&=& \phib_\mbf - \bar\theta^+ \psib_{+,\mbf}~, \\
{\bar\Xi}_\mbf &=& \psib_{-,\mbf} - \theta^- i \Delta_z \phib_\mbf-  \bar\theta^+ \Fb_\mbf + i  \theta^-   \bar\theta^+  \Delta_z \psib_{+,\mbf}~.  \nonumber 
\eeq

The $Q_-$ and $\bar{Q}_+$ transformations of (\ref{latticesuperfieldAppx}) are easily read off by noting that they are simply given by shifts of $\theta^-$ and $\bar\theta^+$ as implied by (\ref{22operator}) and (\ref{Aconjugation}). The transformations under the rest of the supercharges are easily worked out starting from eqn.~(\ref{22operator}) and paying attention to the change of basis, with the result, denoting $\delta^\prime = \epsilon^+ Q_+ - \bar\epsilon^- \bar{Q}_-$:  
\beq
\label{extrasusy}
\delta^\prime \phi_\mbf &=& \epsilon^+ \psi_{+,\mbf},~~~ \delta^\prime \psi_{-,\mbf} = - \epsilon^+ F_\mbf - i \; \bar\epsilon^- \Delta_z \phi_\mbf, \nonumber\\
 \delta^\prime \psi_{+,\mbf} &=& 0,~~~ ~~~~~~~~~~~ \delta^\prime F_\mbf = - i \; \bar\epsilon^- \Delta_z \psi_{+,\mbf}~.
\eeq
The $\delta^\prime$ transformations on the components if $\bar{U}$ and $\bar{\Xi}$ 
can be found by an $I$-conjugation of (\ref{extrasusy}), using the 
rules of (\ref{involution}) and, $(\epsilon^+)^I = i \bar{\epsilon}^-$, 
$(\bar{\epsilon}^-)^I = i \epsilon^+$. It is also easily checked that 
the transformations generated by $Q_\pm$ and $\bar{Q}_\pm$ defined 
above obey a discretized version of the continuum algebra with 
$2\partial_+$ replaced by $\Delta_{\bar{z}}$ and $2\partial_-$ 
by $- \Delta_z$; and that the free action, including the 
$F$-type Wilson term (discussed in section \ref{lift2}) is invariant 
under all four supersymmetries, as in the 1d case.

\section{Twisted mass and Wilson terms}
\label{lift1}
In this section, we continue our study of the nonlinear sigma model. 
In order to save space, we will only consider the case of the $CP^1$ model, as we believe that this case is sufficiently general to allow us to make all our points; needless to say, our results are more generally valid.

In projective coordinates, the bosonic part of the  $CP^1$ model is  described by a single complex scalar field $\phi$ with K\" ahler metric $
K_{\phi\phib} \sim (1 + \phib \phi)^{-2}$ 
(in this section, we will not worry about normalization  unless absolutely necessary).
In the supersymmetric case, $\phi$ is the lowest component of a chiral superfield. 
In its supersymmetric lattice version the model is described by a single set of 
lattice superfields $U, \Xi$, and their I-conjugates, with component action  (\ref{latticeDtermALL}),  and with $K_{\phi \phib}$ substituted for the general $K_{I \bar{J}}$, i.e.:
\beq
\label{cp1original}
S_{CP^1} = a^2 \sum_{ \mbf }  \int d \bar\theta^+ d \theta^- { \bar\Xi_\mbf    \; \Xi_\mbf \over ( 1 + {\bar{U}}_\mbf \;   U_\mbf )^2 }~.
\eeq

The continuum $CP^1$ model has three isometries, representing the   action of the $SO(3)$ global symmetry on $S^2 \simeq CP^1$ in   projective coordinates. They are generated by the three holomorphic Killing vectors:\footnote{See, for example \cite{WessBagger}.}
\beq
\label{killing1}
X^1(\phi) = - {i \over 2}(1 - \phi^2)~, ~~X^2(\phi) = {1 \over 2} (1 + \phi^2)~, ~~ X^3(\phi) = - i \phi~, 
\eeq
under which the component fields transform as follows:
\beq
\label{cp1isometry2}
\phi  \rightarrow \phi  + \alpha_i X^i  ~, \phib \rightarrow \phib + \alpha_i \bar{X}^i  ~, 
\psi_{\pm}  \rightarrow \psi_\pm  + \alpha_i {\partial X^i \over \partial \phi} \psi_\pm~, \psib_\pm \rightarrow \psib_\pm  + \alpha_i  {\partial \bar{X}^i \over \partial \bar{\phi} } \psib_\pm~,
\eeq
where $\alpha^i, i = 1,2,3$ are real parameters.
As eqns.~(\ref{killing1}), (\ref{cp1isometry2}) show,  only one of the isometries, generated by the $X^3$ Killing vector (an $SO(2) \subset  SO(3)$ rotation), acts linearly:
\beq
\label{cp1isometry}
\phi \rightarrow e^{- i \alpha_3} \phi, \psi_\pm \rightarrow e^{-i \alpha_3} \psi_\pm, ~ \phib \rightarrow e^{  i \alpha_3 }  \phib ~, \psib_\pm \rightarrow e^{i \alpha_3} \psib_\pm~.
\eeq
In fact, being the only linearly acting isometry,  (\ref{cp1isometry}) is the only  global symmetry exactly  preserved on the lattice---the $SO(3)/SO(2)$   isometries  fail, just like supersymmetry of the  interaction terms, because of the failure of the Leibnitz rule for lattice derivatives, and are only preserved up to powers of the lattice spacing.\footnote{If one uses a real three vector to describe the bosonic sector of the model, the $SO(3)$ global symmetry can be exactly preserved on the lattice; however, lattice supersymmetry is explicitly broken; see \cite{DiVecchia:1983ax} (we note that fermion doubling and anomalies were not considered there).}

If we now eliminate the auxiliary field $F$ from eqn.~(\ref{latticeDtermALL}), via its equation of motion, we arrive at a $Q_-, \bar{Q}_+$ invariant partition function with measure:
\beq
\label{cp1measure}
\prod_{\mbf} d \phi d \phib \;  (1 + \phi \phib)^2 \; d \psi_+ d \psib_+ d \psi_- d \psib_-  ~.\eeq
It is important that the  measure (\ref{cp1measure}) is invariant under the entire $SO(3)$ isometry group  (with action given in (\ref{cp1isometry2}), which implies that $d \psi_\pm  \rightarrow d \psi_\pm/(1 + \alpha_i \partial X^i/\partial \phi)$, etc.); the presence of the $(1 +  \phi \phib)^2$ factor in the measure is crucial to ensure this invariance. 
The  ``predecessor" of (\ref{cp1measure}),   eqn.~(\ref{measure22}), is invariant 
under general local holomorphic field redefinitions.  Note also that the seemingly divergent integral over  the zero modes of $\phi, \phib$ due to the measure factor in (\ref{cp1measure}) is compensated by the curvature factor, $\sim (1 + \phib \phi)^{-4}$,  coming from the fermion zero mode integration. 
Thus, the model with action (\ref{latticeDtermALL}), with $F$ eliminated, and measure given by (\ref{cp1measure}), appears to define a 
finite partition function. Nevertheless, as explained in the end of  section \ref{22susysigma}, it can not have the desired continuum limit: an examination of the perturbative spectrum of (\ref{latticeDtermALL}) reveals the presence of doublers, which ultimately lead to  the vanishing of the $U(1)_A$ anomaly. 

In this section, we will attempt to lift the doublers by employing a 
lattice version of a known supersymmetric deformation of the 
continuum theory---the addition of twisted mass terms; this method
has previously been applied in \cite{Catterall:2003uf}. 
It  amounts to first gauging a holomorphic isometry by introducing a corresponding vector superfield $V$, then giving an expectation value to the scalars in the vector multiplet, and  subsequently decoupling all its fluctuations (this is described in superspace in \cite{Gates:1983py}).  The vevs of the scalars in $V$ lead   to the appearance of a central charge in the supersymmetry algebra. As discussed above, on the lattice the only exact isometry we 
have at our disposal  is the one generated by $X^3$ (\ref{cp1isometry}). 
We can describe the introduction of a  twisted mass term in our superspace 
formalism by means of  a spurion field ${\cal{V}} = \bar\theta^+ \theta^- i z$, with $z$ assumed
to be a real constant (so that $I$ maps ${\cal{V}} \rightarrow {\cal{V}}$),  into (\ref{Dsuperspace}), restricted to the $CP^1$ case:
\beq
\label{cp1twisted1}
S_{twisted}^{(1)} =  a^2 \sum_{ \mbf }  \int d \bar\theta^+ d \theta^- { \bar\Xi_\mbf \; e^{\cal{V}}  \; \Xi_\mbf \over ( 1 + {\bar{U}}_\mbf \; e^{\cal{V}}  \; U_\mbf )^2 } ~.
\eeq
The introduction of the explicit $\bar\theta^+ \theta^-$ into the superspace action breaks the $\theta$-shift  symmetry; however, supersymmetry can be redefined by supplementing shifts of $\theta$ by appropriate field transformations. More precisely, the $\hat{\delta}$ variation  (\ref{22operator}) of ${\cal{V}}$ is $\hat{\delta} e^{\cal{V}} = i z \bar\epsilon^+ \theta^-  + i z \bar\theta^+ \epsilon^-$. Thus, to restore ``supersymmetry," we are forced to absorb $\hat{\delta} e^{\cal{V}}$ by    redefining the action of supersymmetry, eqn.~(\ref{22operator}),  $\hat{\delta} \rightarrow \hat{\delta}^\prime$ on the fields, in a manner consistent with the action of the involution:
\beq
\label{twistedsusy}
\hat{\delta}^\prime U &=&\hat{\delta} U  - i z \bar\epsilon^+ \theta^- U~, ~~~ \hat{\delta}^\prime \Xi =  \hat{\delta} \Xi - i z \bar\epsilon^+ \theta^- \Xi,  \nonumber \\
 \hat{\delta}^\prime \bar{U} &=&\hat{\delta} \bar{U} - i z \bar\theta^+ \epsilon^- \bar{U}~, ~~~ \hat{\delta}^\prime \bar\Xi = \hat{\delta} \bar\Xi  - i z \bar\theta^+ \epsilon^- \bar\Xi~.
\eeq
The corresponding modification of the  component field transformations can be immediately read off from 
(\ref{twistedsusy}). In particular, one observes that only $\psi_-$, $\psib_+$, $F$, and $\bar{F}$ experience an extra $z$- and $\epsilon$- dependent shift and that the measure (\ref{measure22}) is invariant.   It is straightforward to compute the anticommutator of the modified supersymmetry acting on the component fields and verify that (\ref{twistedsusy}) realizes the algebra with central charge, $\{ \bar{Q}_+, Q_-\} = 2 i z$.

The effect of introducing the spurion into  (\ref{cp1twisted1}) is easy to read off; for example, at the bilinear level, we see that (\ref{cp1twisted1}) gives a $U(1)_A$ violating mass term $\sim z \psib_- \psi_+$ to the fermions, but not to the scalars. 
Since  introducing ${\cal{V}} = \bar\theta^+ \theta^- i z$ breaks $U(1)_A$ but preserves all other symmetries, we have to revisit our  arguments that lead to establishing eqn.~(\ref{Dsuperspace}) as the most general lattice action consistent with the symmetries. The lattice supersymmetry as well as dimensional arguments (recall that $U$ is dimensionless, while $z$ has mass dimension 1), now allow   general $U(1)_A$--violating, but $U(1)_V$, $I$, and $Z_4$ preserving terms of the form: 
\beq
\label{ua1violatingtwisted}
S_{twisted}^{(2)} = - i z a^2 \sum_{\mbf} \int d \bar\theta^+ d \theta^-   \; f( {\bar{U}}_\mbf   e^{\cal{V}}   U_\mbf  )~.
\eeq
For the $CP^1$ model, the choice  $f(x) = {x \over (1 + x)^{2}}$  yields, in the naive continuum limit, the continuum theory result  for an  $SO(2)$ twisted mass term. In addition to the $\sim z \psib_- \psi_+$ mass term from (\ref{cp1twisted1})   one finds that (\ref{ua1violatingtwisted})  gives rise to additional fermion    $\sim z \psib_+ \psi_-$ as well as  scalar $\sim z^2 \phib \phi$ mass terms. The rest of the twisted model lagrangian can be easily worked out by expanding  $S_{twisted}^{(1)} + S_{twisted}^{(2)}$ in components (with the $f$ given just below (\ref{ua1violatingtwisted})). The naive continuum limit of the component lattice action is easily seen to coincide with the general formulae of \cite{Gates:1983py}.

Let us now ask whether   twisted mass terms  give an opportunity to lift the doublers while preserving the exact lattice supersymmetry. We start by  introducing nonlocal twisted mass terms 
by inserting a lattice laplacian in the numerators of  (\ref{cp1twisted1}), (\ref{ua1violatingtwisted}):
\beq
\label{twistednonlocalmass1}
S_r &=&
 a^2 \sum_{\mbf} \int d \bar\theta^+ d \theta^-  \times  \\
&\times&r a^2 ~  { \bar\Xi_\mbf e^{\cal{V}} \Delta^2 \Xi_\mbf + \Delta^2 \bar\Xi_\mbf e^{\cal{V}}   \Xi_\mbf- i z \; {\bar{U}}_\mbf e^{\cal{V}} \Delta^2 U_\mbf - i z \; \Delta^2 {\bar{U}}_\mbf e^{\cal{V}}  U_\mbf \over \left(1 + {\bar{U}}_\mbf e^{\cal{V}} U_\mbf \right)^2 } ~.\nonumber 
\eeq
When added to the original action, (\ref{cp1original}),  it 
is easy to check that (\ref{twistednonlocalmass1}), upon 
taking the twisted nonlocal mass to scale (on dimensional grounds) 
as $z \sim 1/a$,  gives mass of order $r z \sim r/a$ to the fermion doublers and of order $\sqrt{|r|} z \sim \sqrt{|r|}/a$ to the bosonic doublers. 
The different scaling with respect to~$r$ for bosons and fermions already follows from the linearity of  $S_r$  in $r$ and its being a mass term for fermions and mass squared term for the scalars (the auxiliary field is irrelevant for the scalar mass, in contrast with the case of superpotential Wilson term of section \ref{lift2}). Besides removing the doublers, 
 the ``Wilson action"  $S_r$ violates the $U(1)_A$ symmetry of (\ref{cp1original}); one expects, then, that the $U(1)_A$ anomaly will be correctly reproduced in the continuum limit, as in the last reference in \cite{Nielsen:1980rz}.

The drawback of using the nonlocal twisted mass terms  (\ref{twistednonlocalmass1}) to eliminate the doublers is the explicit breaking of the lattice supersymmetry (as witnessed already by the different masses of the bosonic and fermionic doublers quoted above). The two terms in the action, (\ref{cp1original}) and (\ref{twistednonlocalmass1}), respect different supersymmetries: the original one (\ref{22operator}) and the modified (\ref{twistedsusy}), respectively. 

One could attempt to reconcile the two supersymmetries by also modifying the leading term by including an $e^{\cal{V}}$ 
 factor there, so that both terms now respect the same twisted supersymmetry,  i.e., by considering the total action:
\beq
\label{totalaction}
S = S_{twisted}^{(1)}  + S_{twisted}^{(2)}+ S_r~.
\eeq
The problem with this approach is that while the twisted lattice supersymmetry is respected, 
 the leading term gives an additional twisted local mass  $\sim z$. Then, a study of the dispersion relation shows that there is no sensible limit of parameters  allowing the doublers to decouple.

To summarize this section: it is possible to use  twisted nonlocal mass 
term deformations of the nonlinear sigma model lattice action to remove 
the doublers and break the anomalous axial symmetry. However, the lattice 
supersymmetry is then explicitly broken. It is not immediately obvious 
whether the supersymmetric continuum limit can be achieved without fine tuning. 
A generic 2d $(2,2)$ nonlinear sigma model is a renormalizable theory (rather 
then superrenormalizable, or even finite, as is the LG model) and logarithmic divergences are present at any order of perturbation theory.  
Whether there exist models where the desired continuum limit (using the twisted nonlocal mass terms)  can 
be achieved without fine tuning deserves further study.

\section{Power counting and finiteness of the supersymmetric lattice LG model} 
\label{graphs}
Here we consider naive power-counting to determine the
superficially divergent subdiagrams that occur in
lattice perturbation theory.  In this exercise, we
follow Reisz~\cite{Reisz:1987da}.  

As alluded to in the main text, we can incorporate the
Wilson terms of \myref{Fsuperspace} directly into the
superpotential action \myref{Fsuperspace1}.  This is done
by a modification of the superpotential \myref{wU} as follows
(note that we restrict to the simplest sort of superpotential 
interaction---cubic---and
that some rescaling of coefficients has been performed):
\beq
W(U) = \sum_\mbf \[ \frac{m}{2} U_\mbf^2 + \frac{g}{3!} U^3_\mbf \]
- \frac{ra}{4} \sum_{\mbf, \nbf} U_\mbf \Delta_{\mbf,\nbf}^2 U_\nbf~.
\eeq
Then one defines
$W'_\mbf = \p W / \p U_\mbf$ and 
$W''_{\mbf,\nbf} = \p^2 W / \p U_\mbf \p U_\nbf$.
This moves the $r$-dependent terms of \myref{latticeFtermlocal}
into the coefficients denoted there as $W_I$ and $W_{IJ}$.

Note that in this formula and all others in our power-counting
considerations, the lattice spacing $a$ is included in the
definition of finite difference operators.  For example,
the lattice laplacian is:
\beq
\Delta_{\mbf,\nbf}^2 = \frac{1}{a^2} \sum_{\mu=1,2}
\( \delta_{\mbf + \hat \mu, \nbf} + \delta_{\mbf - \hat \mu, \nbf}
- 2 \delta_{\mbf, \nbf} \)~.
\label{llap}
\eeq
Following the construction outlined in section \ref{22section}, we obtain the action:
\beq
a^{-2} S &=& -i \psib_- \Delta_\zb \psi_- + i \psi_+ \Delta_z \psib_+
-  \phib \Delta_z \Delta_\zb \phi - \Fb F \nnn
&& - \psib_- \bbar{W}'' \psib_+ - \psi_+ W'' \psi_-
+ W'(\phi) (F + i \Delta _\zb \phi)
+ \bbar{W}'(\phib) (\Fb - i \Delta_z \phib)~,
\eeq
Integrating out $F,\Fb$, the action becomes:
\beq
a^{-2} S = [\Delta_z \phib + i W'(\phi) ]_\mbf
[\Delta_\zb \phi - i \bbar{W}'(\phib) ]_\mbf
+ \chib_\mbf M_{\mbf,\nbf}(\phi,\phib) \chi_\nbf~,
\eeq
where the fermions have been organized according to:
\beq
\chib = (\psi_+, \psib_-), \qquad
\chi = \binom{\psi_-}{\psib_+}, \qquad
M = \begin{pmatrix} -W''(\phi) & i \Delta_z \cr
-i \Delta_\zb & - \bbar{W}''(\phib) \end{pmatrix}
\label{fmat}~.
\eeq
The action can be written as $S=S_0 + S_{int}$,
where $S_0$ is quadratic in fields.  Explicitly,
\beq
a^{-2} S_0 &=& \phib \[ -  \Delta_z \Delta_\zb + 
(\bar m - \frac{\bar ra}{2} \Delta^2 )
(m - \frac{ra}{2} \Delta^2 ) \] \phi
+ \chib M_0 \chi ~,\nnn~
M_0 &=& \begin{pmatrix} -(m-\frac{ra}{2} \Delta^2) & i \Delta_z \cr
-i \Delta_\zb & -(\bar m - \frac{\bar ra}{2} \Delta^2 )  ~
\end{pmatrix}~.
\eeq
For the interaction terms, it is convenient to introduce
chirality projection operators for the fermions:
$L = \half(1 + \s_3)$ and $R = \half(1- \s_3)$.  Then:
\beq
a^{-2} S_{int} &=& - g \phi \chib L \chi - \bar g \phib \chib R \chi
+ \half \bar g m \phib^2 \phi + \half g \bar m \phi^2 \phib 
+ \fourth |g|^2 \phi^2 \phib^2 \nnn &&
- \fourth ra \bar g \phib^2 \Delta^2 \phi
- \fourth \bar ra g \phi^2 \Delta^2 \phib
+ \frac{i}{2} g \phi^2 \Delta_\zb \phi - \frac{i}{2} \bar g \phib^2 \Delta_z \phib~.
\label{uwre}
\eeq

Each loop integral has
associated with it
$
\frac{1}{(Na)^2} \sum_{\kbf} \to \int \frac{d^2 k}{(2 \pi)^2}
$.
Since this scales like $a^{-2}$, the degree of
divergence $D=2$, just as in the continuum theory.
$E_{B,F}$ are the number of external
boson and fermion lines resp.  $I_{B,F}$ count internal lines.
The boson propagator is given by:
\beq
\tilde G_\kbf = \[ a^{-2} \sum_\mu s^2(k_\mu) + |m(\kbf)|^2 \]^{-1}~,
\eeq
where here and below 
$s(k_\mu) = \sin (2\pi k_\mu/N)$, $c(k_\mu) = \cos (2\pi k_\mu/N)$
and
\beq
m(\kbf) = m + 2ra^{-1} \sum_\mu s^2(k_\mu/2), \qquad m_s(\kbf) = \real m(\kbf),
\qquad m_p(\kbf) = \imag m(\kbf)~.
\eeq
Since $\tilde G_\kbf \sim a^2$, it  thus has $D= -2$, just as in
the continuum.  The Wilson mass term does not alter the degree
of divergence, since what is important is the dependence on $a$.
Similarly, the fermion propagator ($\gamma_3=\s_3, \; 
\gamma_1 = -\s_2, \; \gamma_2 = \s_1$),
\beq
\tilde D_\kbf = \tilde G_\kbf^{-1} \( m_s(\kbf) - i \gamma_3 m_p(\kbf) 
- i a^{-1} \sum_\mu s(k_\mu) \gamma_\mu \)~,
\label{jasr}
\eeq
has $D= -1$, just as in the continuum.

The power counting for the vertices on the  lattice, however, is different
from the continuum.  Below,
$V_1$ is the number of $\phi^2 \phib$ vertices,
$V_2$ is the number of $\phi^2 \phib^2$ vertices,
$V_3$ is the number of $\phi \chib L \chi$ vertices, and
$V_4$ is the number of $\phi^2  \Delta_{\zb} \phi$ vertices.
Conjugate vertices are also counted in these quantities.
In contrast to the continuum, the vertices counted by 
$V_1$ and $V_4$ carry degree of divergence $D=1$
because they scale like $a^{-1}$.  For example, the $\phi \phib^2$ vertices
arise from:
\beq
S_{int} \ni a^2 \bar g \phib^2 \( \half m \phi - \fourth ra \Delta^2 \) \phi~.
\eeq
Since $a \Delta^2 \sim a^{-1}$, the interaction associated
with supersymmetrization of the Wilson mass term yields $D=1$
for the corresponding vertex.

Taking these contributions to $D$ for a given diagram
into account, as well as the usual constraints on lines
and vertices, we obtain ($L$ is the number of loops):
\beq
D&=&2L-I_F-2I_B+V_1+V_4~, \nnn
L&=&I_B+I_F-V_1-V_2-V_3-V_4+1~, \nnn
E_B+2I_B&=&3V_1+4V_2+V_3+3V_4~, \\
E_F+2I_F&=&2V_3~. \nonumber
\eeq
We can eliminate $I_{B,F}$ and $L$ to obtain:
\beq
D=2-\half E_F-V_1-2V_2-V_3-V_4~.
\eeq
Taking into account constraints that arise from
the Feynman rules, there exist 7 superficially
divergent types of subdiagrams,
consisting of tadpoles and 2-point functions; see Fig.~\ref{fig1}.

A straightforward application of \myref{uwre}-\myref{jasr}
shows that the net result
for a particular choice of external lines either
vanishes or is finite.  In the case of the 1-point
function, the two diagrams cancel exactly.
In the case of the 2-point functions, $D=0$.  Hence it suffices
to check at external momentum $\kbf_{ext}=0$.  This
is because $d/d \kbf \sim a$, so that contributions
at nonzero $\kbf$ are suppressed by further powers of $a$.
The diagrams associated with $\vev{\phi \phi}$ cancel
exactly at any $\kbf_{ext}$.  
The cancellations leading to these results are a
consequence of the one exact supersymmetry.

The diagrams associated with $\vev{\phi \phib}$ sum to a nonzero
but finite quantity, thus deserving special mention.  The $D=0$
parts cancel exactly, again due to the exact lattice supersymmetry.
What is left is only the $D= -2$ part.
As shown by Reisz, $D < 0$
contributions to lattice perturbation theory necessarily
approach their continuum values in the $a \to 0$ limit \cite{Reisz:1987da}. 
It follows that lattice perturbation theory using the $Q_A$-exact
action is finite and reproduces the results of the
continuum perturbation theory.  This is an essential feature in
rendering the $\ord{a}$ irrelevant operators harmless
in the $a \to 0$ limit.  

\FIGURE{ 
\includegraphics[height=4in,width=5.5in,bb=65 400 500 750,clip]{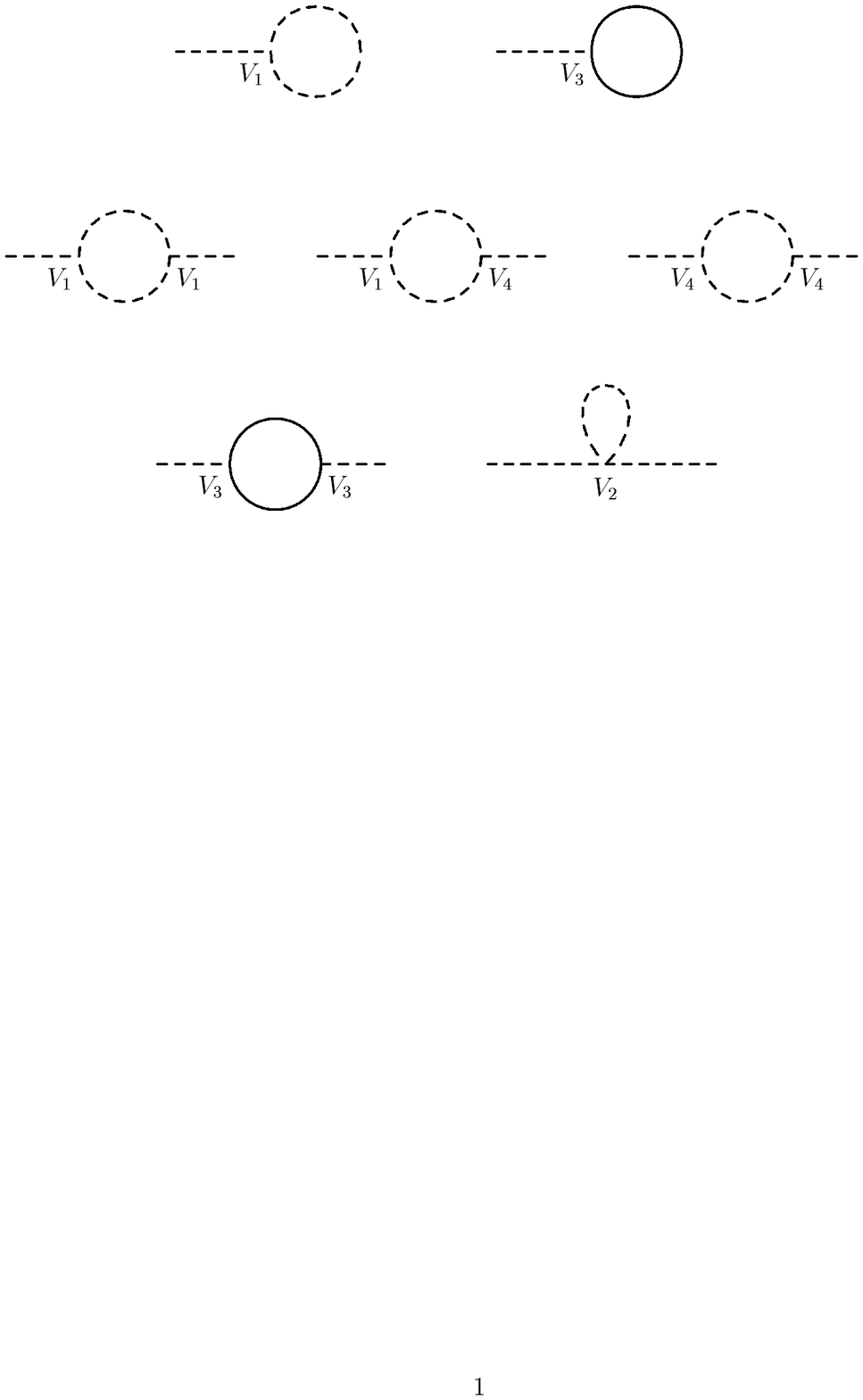}
\caption{Types of superficially divergent diagrams.
Dashed lines are bosons and solid lines are fermions. 
\label{fig1}}
}

Many of these cancellations arise due to the Ward identities 
that follow from the exact $Q_A$ supersymmetry, which also 
restricts finite renormalizations of the lattice action.  For example, since
$Q_A \psi_+ = F+i\Delta_{\zb} \phi$, and since (as in quantum mechanics, see the end of appendix A)  correlation functions of $Q$-exact quantities vanish, we find that
\beq
\vev{F} = -i\vev{\Delta_{\zb} \phi}
\eeq
exactly on the lattice. Translational symmetry then implies that the r.h.s.~vanishes, hence 
$F$ tadpoles are forbidden (at the component loop level discussed above, this Ward identity implies that vanishing $\phi$ tadpoles imply vanishing of the $\phi^2$ graphs at zero momentum and vice versa).

We should also note that  the consequences of the exact nilpotent $Q_A$ supersymmetry of the partition function transcend perturbation theory. The nilpotent supersymmetry of the partition function implies localization (see, e.g., the second reference in \cite{Hori:2000kt}). It should then be possible, 
through a careful application of localization, to study the  Witten index for the LG models on the lattice, along the lines discussed in appendix A. We leave this for future work.

As far as the practical problems of simulations are concerned, at this point we also note that the fermion matrix \myref{fmat} can be cast in a real basis, through the change of coordinates:\footnote{We
thank S.~Catterall for pointing this out to us.}
\beq
&& \chi_1 = \frac{1}{\sqtw} (\eta_1 + i \eta_2), \qquad
\chi_2 = \frac{1}{\sqtw} (\eta_1 - i \eta_2), \nnn
&& \chib_1 = \frac{1}{\sqtw} (\bar \eta_1 - i \bar \eta_2), \qquad
\chib_2 = \frac{1}{\sqtw} (\bar \eta_1 + i \bar \eta_2)~.
\eeq
The matrix then becomes:
\beq
M =
\begin{pmatrix}
-\real W'' + \Delta_2 & \imag W'' + \Delta_1 \cr
-\imag W'' + \Delta_1 & -\real W'' - \Delta_2
\end{pmatrix}~.
\eeq
For a square $N \times N$ lattice, $\det M > 0$, because
the eigenvalues always come in complex conjugate pairs.
However, for a more general $N_1 \times N_2$ lattice,
we find that unpaired real eigenvalues can occur.  These
then allow for $\det M$ to be both positive or negative;
however, in random Gaussian samples of boson configurations
we find that $\det M < 0$ occurs only rarely.
For purposes of Monte Carlo simulation, the square
lattices are much to be preferred, since they do not suffer
from a sign problem.\footnote{Many other supersymmetric
lattice theories suffer from a sign problem.  For example,
the $(1,1)$ Wess-Zumino model studied in \cite{Catterall:2003ae}, or
the 2d super-Yang-Mills models based on 
deconstruction \cite{Giedt:2004tn}.}  
These are the type of lattices
studied in \cite{Catterall:2001fr}.  They allow for a
faithful representation of the fermion determinant through
real pseudofermions $y$:
\beq
S_{p.f.} = \half y_\mbf [(M^T M)^{-1}]_{\mbf,\nbf} y_\nbf
\eeq
since $[\det (M^T M)]^{1/2} = \det M$ if $\det M$ is always
positive.  In particular, this allows for a hybrid Monte
Carlo simulation, as was performed in \cite{Catterall:2001fr}.
In practice, this is an advantage since it avoids
the systematic errors of, say, the $R$-algorithm.


\begin{thebibliography}{99}

\bibitem{Wess:1973kz}
J.~Wess and B.~Zumino,
``A Lagrangian Model Invariant Under Supergauge Transformations,''
Phys.\ Lett.\ B {\bf 49} (1974) 52.

\bibitem{Iliopoulos:1974zv}
J.~Iliopoulos and B.~Zumino,
``Broken Supergauge Symmetry And Renormalization,''
Nucl.\ Phys.\ B {\bf 76} (1974) 310.

\bibitem{Seiberg:1993vc}
N.~Seiberg,
``Naturalness versus supersymmetric nonrenormalization theorems,''
Phys.\ Lett.\ B {\bf 318} (1993) 469
[arXiv:hep-ph/9309335].

\bibitem{ADS}
I.~Affleck, M.~Dine and N.~Seiberg,
I.~Affleck, M.~Dine and N.~Seiberg,
``Supersymmetry Breaking By Instantons,''
Phys.\ Rev.\ Lett.\  {\bf 51} (1983) 1026;
``Dynamical Supersymmetry Breaking In Supersymmetric QCD,''
Nucl.\ Phys.\ B {\bf 241} (1984) 493;
I.~Affleck, M.~Dine and N.~Seiberg,
``Dynamical Supersymmetry Breaking In Four-Dimensions And Its Phenomenological
Implications,''
Nucl.\ Phys.\ B {\bf 256} (1985) 557.


\bibitem{Kaplan:2002wv}
D.~B.~Kaplan, E.~Katz and M.~Unsal,
``Supersymmetry on a spatial lattice,''
JHEP {\bf 0305} (2003) 037
[arXiv:hep-lat/0206019];
A.~G.~Cohen, D.~B.~Kaplan, E.~Katz and M.~Unsal,
``Supersymmetry on a Euclidean spacetime lattice. I: A target theory with  four
supercharges,''
JHEP {\bf 0308} (2003) 024
[arXiv:hep-lat/0302017];
A.~G.~Cohen, D.~B.~Kaplan, E.~Katz and M.~Unsal,
``Supersymmetry on a Euclidean spacetime lattice. II: Target theories with
eight supercharges,''
JHEP {\bf 0312} (2003) 031
[arXiv:hep-lat/0307012].


\bibitem{Sugino:2003yb}
F.~Sugino,
``A lattice formulation of super Yang-Mills theories with exact
supersymmetry,''
JHEP {\bf 0401} (2004) 015
[arXiv:hep-lat/0311021];
F.~Sugino,
``Super Yang-Mills theories on the two-dimensional lattice with exact
supersymmetry,''
JHEP {\bf 0403} (2004) 067
[arXiv:hep-lat/0401017].


\bibitem{Curci:1986sm}
G.~Curci and G.~Veneziano,
``Supersymmetry And The Lattice: A Reconciliation?,''
Nucl.\ Phys.\ B {\bf 292} (1987) 555.


\bibitem{Kaplan:1999jn}
D.~B.~Kaplan and M.~Schmaltz,
``Supersymmetric Yang-Mills theories from domain wall fermions,''
Chin.\ J.\ Phys.\  {\bf 38}, 543 (2000)
[arXiv:hep-lat/0002030];
G.~T.~Fleming, J.~B.~Kogut and P.~M.~Vranas,
``Super Yang-Mills on the lattice with domain wall fermions,''
Phys.\ Rev.\ D {\bf 64} (2001) 034510
[arXiv:hep-lat/0008009];
N.~Maru and J.~Nishimura,
``Lattice formulation of supersymmetric Yang-Mills theories without  fine-tuning,''
Int.\ J.\ Mod.\ Phys.\ A {\bf 13} (1998) 2841
[arXiv:hep-th/9705152].

\bibitem{Elitzur:1982vh}
S.~Elitzur, E.~Rabinovici and A.~Schwimmer,
``Supersymmetric Models On The Lattice,''
Phys.\ Lett.\ B {\bf 119} (1982) 165.

\bibitem{Elitzur:1983nj}
S.~Elitzur and A.~Schwimmer,
``N=2 Two-Dimensional Wess-Zumino Model On The Lattice,''
Nucl.\ Phys.\ B {\bf 226}, 109 (1983).

\bibitem{Cecotti:1982ad}
S.~Cecotti and L.~Girardello,
``Stochastic Processes In Lattice (Extended) Supersymmetry,''
Nucl.\ Phys.\ B {\bf 226}, 417 (1983).

\bibitem{Sakai:1983dg}
N.~Sakai and M.~Sakamoto,
``Lattice Supersymmetry And The Nicolai Mapping,''
Nucl.\ Phys.\ B {\bf 229}, 173 (1983).

\bibitem{Catterall:2000rv}
S.~Catterall and E.~Gregory,
``A lattice path integral for supersymmetric quantum mechanics,''
Phys.\ Lett.\ B {\bf 487} (2000) 349
[arXiv:hep-lat/0006013].

\bibitem{Catterall:2001wx}
S.~Catterall and S.~Karamov,
``A two-dimensional lattice model with exact supersymmetry,''
Nucl.\ Phys.\ Proc.\ Suppl.\  {\bf 106} (2002) 935
[arXiv:hep-lat/0110071].

\bibitem{Catterall:2001fr}
S.~Catterall and S.~Karamov,
``Exact lattice supersymmetry: the two-dimensional N = 2 Wess-Zumino  model,''
Phys.\ Rev.\ D {\bf 65} (2002) 094501
[arXiv:hep-lat/0108024].

\bibitem{Catterall:2003wd}
S.~Catterall,
``Lattice supersymmetry and topological field theory,''
JHEP {\bf 0305} (2003) 038
[arXiv:hep-lat/0301028];
S.~Catterall,
``Lattice supersymmetry and topological field theory,''
Nucl.\ Phys.\ Proc.\ Suppl.\  {\bf 129} (2004) 871
[arXiv:hep-lat/0309040].

\bibitem{Catterall:2003uf}
S.~Catterall and S.~Ghadab,
``Lattice sigma models with exact supersymmetry,''
JHEP {\bf 0405} (2004) 044
[arXiv:hep-lat/0311042].

\bibitem{Golterman:1988ta}
M.~F.~L.~Golterman and D.~N.~Petcher,
``A Local Interactive Lattice Model With Supersymmetry,''
Nucl.\ Phys.\ B {\bf 319}, 307 (1989).

\bibitem{Dondi:1976tx}
P.~H.~Dondi and H.~Nicolai,
``Lattice Supersymmetry,''
Nuovo Cim.\ A {\bf 41}, 1 (1977).

\bibitem{Lunin:1999ib}
O.~Lunin and S.~Pinsky,
``SDLCQ: Supersymmetric discrete light cone quantization,''
AIP Conf.\ Proc.\  {\bf 494}, 140 (1999)
[arXiv:hep-th/9910222]; 
for a more recent list of references, see, e.g.,
M.~Harada, J.~R.~Hiller, S.~Pinsky and N.~Salwen,
``Improved results for N = (2,2) super Yang-Mills theory using supersymmetric
discrete light-cone quantization,''
[arXiv:hep-th/0404123].


\bibitem{Beccaria:2004ds}
M.~Beccaria, G.~F.~De Angelis, M.~Campostrini and A.~Feo,
``Phase diagram of the lattice Wess-Zumino model from rigorous lower bounds on
the energy,''
arXiv:hep-lat/0405016, and references therein.

\bibitem{Kikukawa:2002as}
Y.~Kikukawa and Y.~Nakayama,
``Nicolai mapping vs. exact chiral symmetry on the lattice,''
Phys.\ Rev.\ D {\bf 66}, 094508 (2002)
[arXiv:hep-lat/0207013].

\bibitem{Hori:2000kt}
K.~Hori and C.~Vafa,
``Mirror symmetry,''
arXiv:hep-th/0002222; see also K. Hori et al., ``Mirror symmetry," Clay
Mathematics Monographs (American Mathematical Society, 2003).


\bibitem{Nielsen:1980rz}
H.~B.~Nielsen and M.~Ninomiya,
``Absence Of Neutrinos On A Lattice. 1. Proof By Homotopy Theory,''
Nucl.\ Phys.\ B {\bf 185} (1981) 20
[Erratum-ibid.\ B {\bf 195} (1982) 541];
``Absence Of Neutrinos On A Lattice. 2. Intuitive Topological Proof,''
Nucl.\ Phys.\ B {\bf 193} (1981) 173;
L.~H.~Karsten and J.~Smit,
``Lattice Fermions: Species Doubling, Chiral Invariance, And The Triangle
Anomaly,''
Nucl.\ Phys.\ B {\bf 183} (1981) 103.

\bibitem{Banks:1982ut}
T.~Banks and P.~Windey,
``Supersymmetric Lattice Theories,''
Nucl.\ Phys.\ B {\bf 198}, 226 (1982).


\bibitem{Reisz:1987da}
T.~Reisz,
``A Power Counting Theorem For Feynman Integrals On The Lattice,''
Commun.\ Math.\ Phys.\  {\bf 116} (1988) 81;
``Renormalization Of Feynman Integrals On The Lattice,''
Commun.\ Math.\ Phys.\  {\bf 117} (1988) 79;
T.~Reisz,
``Power Counting And Renormalization In Lattice Field Theory,''
in {\it Proceeding, Nonperturbative Quantum Field Theory,}
Cargese, 1987.

\bibitem{Montvay}I.~Montvay and G.~Munster, ``Quantum Fields on a Lattice," (Cambridge University Press, 1994).

\bibitem{Catterall:2003ae}
S.~Catterall and S.~Karamov,
``A lattice study of the two-dimensional Wess Zumino model,''
Phys.\ Rev.\ D {\bf 68} (2003) 014503
[arXiv:hep-lat/0305002].

\bibitem{Giedt:2004tn}
J.~Giedt,
``Deconstruction, 2d lattice super-Yang-Mills, and the dynamical lattice
spacing,''
arXiv:hep-lat/0405021;
J.~Giedt,
``The fermion determinant in (4,4) 2d lattice super-Yang-Mills,''
Nucl.\ Phys.\ B {\bf 674} (2003) 259
[arXiv:hep-lat/0307024];
J.~Giedt,
``Non-positive fermion determinants in lattice supersymmetry,''
Nucl.\ Phys.\ B {\bf 668} (2003) 138
[arXiv:hep-lat/0304006].

\bibitem{DiVecchia:1983ax}
P.~Di Vecchia, R.~Musto, F.~Nicodemi, R.~Pettorino, P.~Rossi and
P.~Salomonson,
``Explicit Evaluation Of Physical Quantities And Supersymmetry Properties
Of Lattice O(N) Sigma Models At Large N,''
Phys.\ Lett.\ B {\bf 127}, 109 (1983).

\bibitem{WessBagger}
J. Wess and J. Bagger, ``Supersymmetry and Supergravity," (Princeton University Press, Princeton, NJ, 1992).

\bibitem{Gates:1983py}
S.~J.~J.~Gates,
``Superspace Formulation Of New Nonlinear Sigma Models,''
Nucl.\ Phys.\ B {\bf 238}, 349 (1984).

                                                                                

\end{thebibliography}
\end{document}